\documentclass[letter]{aa} 
\usepackage{graphicx}
\usepackage[fleqn]{cases} 
\usepackage{mathtools} 
\usepackage{natbib,twoopt}
\usepackage[colorlinks,linkcolor=black,urlcolor=black,citecolor=black]{hyperref} 
\bibpunct{(}{)}{;}{a}{}{,}             
\makeatletter
  \newcommandtwoopt{\citeads}[3][][]{\href{http://adsabs.harvard.edu/abs/#3}%
    {\def\hyper@linkstart##1##2{}%
     \let\hyper@linkend\@empty\citealp[#1][#2]{#3}}}
  \newcommandtwoopt{\citepads}[3][][]{\href{http://adsabs.harvard.edu/abs/#3}%
    {\def\hyper@linkstart##1##2{}%
     \let\hyper@linkend\@empty\citep[#1][#2]{#3}}}
  \newcommandtwoopt{\citetads}[3][][]{\href{http://adsabs.harvard.edu/abs/#3}%
    {\def\hyper@linkstart##1##2{}%
     \let\hyper@linkend\@empty\citet[#1][#2]{#3}}}
  \newcommandtwoopt{\citeyearads}[3][][]%
    {\href{http://adsabs.harvard.edu/abs/#3}
    {\def\hyper@linkstart##1##2{}%
     \let\hyper@linkend\@empty\citeyear[#1][#2]{#3}}}
\makeatother
\usepackage{txfonts}
\begin{document} 
\title{The slowly pulsating B-star 18\,Peg: A testbed for upper main sequence stellar evolution%
\thanks{Based on observations collected at the European Organisation for Astronomical Research in the Southern Hemisphere under
ESO programmes 265.C-5038(A), 069.C-0263(A), and 073.D-0024(A).
\newline
Based on observations collected at the Centro Astronómico Hispano Alemán (CAHA) at Calar Alto, operated jointly by the Max-Planck Institut f\"ur Astronomie and the Instituto de Astrofísica de Andalucía (CSIC), proposals H2005-2.2-016 and H2015-3.5-008.
\newline
Based on observations made with the William Herschel Telescope operated on the island of La Palma by the Isaac Newton Group in the Spanish Observatorio del Roque de los Muchachos of the Instituto de Astrofísica de Canarias, proposal W15BN015. 
\newline
Based on observations obtained with telescopes of the University Observatory Jena, which is operated by the Astrophysical Institute of the Friedrich-Schiller-University. 
}
}
\author{
A.~Irrgang\inst{\ref{remeis}}
\and
A.~Desphande\inst{\ref{imperial}}
\and
S.~Moehler\inst{\ref{eso}}
\and
M.~Mugrauer\inst{\ref{jena}}
\and
D.~Janousch\inst{\ref{dieterskirchen}}
}
\institute{
Dr.~Karl~Remeis-Observatory \& ECAP, Astronomical Institute, Friedrich-Alexander University Erlangen-N\"urnberg (FAU),\\ Sternwartstr.~7, 96049 Bamberg, Germany\\ \email{andreas.irrgang@fau.de}\label{remeis}
\and
Imperial College London, Blackett Lab, Prince Consort Rd., London SW7 2AZ, United Kingdom\label{imperial}
\and
European Southern Observatory, Karl-Schwarzschild-Str. 2, 85748 Garching, Germany\label{eso}
\and
Astrophysikalisches Institut und Universit\"ats-Sternwarte Jena, Schillerg\"a{\ss}chen 2, 07745 Jena, Germany\label{jena}
\and
Sternwarte Dieterskirchen, Roigerstr. 6, 92542 Dieterskirchen, Germany\label{dieterskirchen}
}
\date{Received 3 May 2016 / Accepted 24 May 2016}

\abstract
{The predicted width of the upper main sequence in stellar evolution models depends on the empirical calibration of the convective overshooting parameter. Despite decades of discussions, its precise value is still unknown and further observational constraints are required to gauge it. Based on a photometric and preliminary asteroseismic analysis, we show that the mid B-type giant 18\,Peg is one of the most evolved members of the rare class of slowly pulsating B-stars and, thus, bears tremendous potential to derive a tight lower limit for the width of the upper main sequence. In addition, 18\,Peg turns out to be part of a single-lined spectroscopic binary system with an eccentric orbit that is greater than 6 years. Further spectroscopic and photometric monitoring and a sophisticated asteroseismic investigation are required to exploit the full potential of this star as a benchmark object for stellar evolution theory.}


\maketitle
%
%
\section{Introduction}
\begin{figure*}
\centering
\includegraphics[width=1\textwidth]{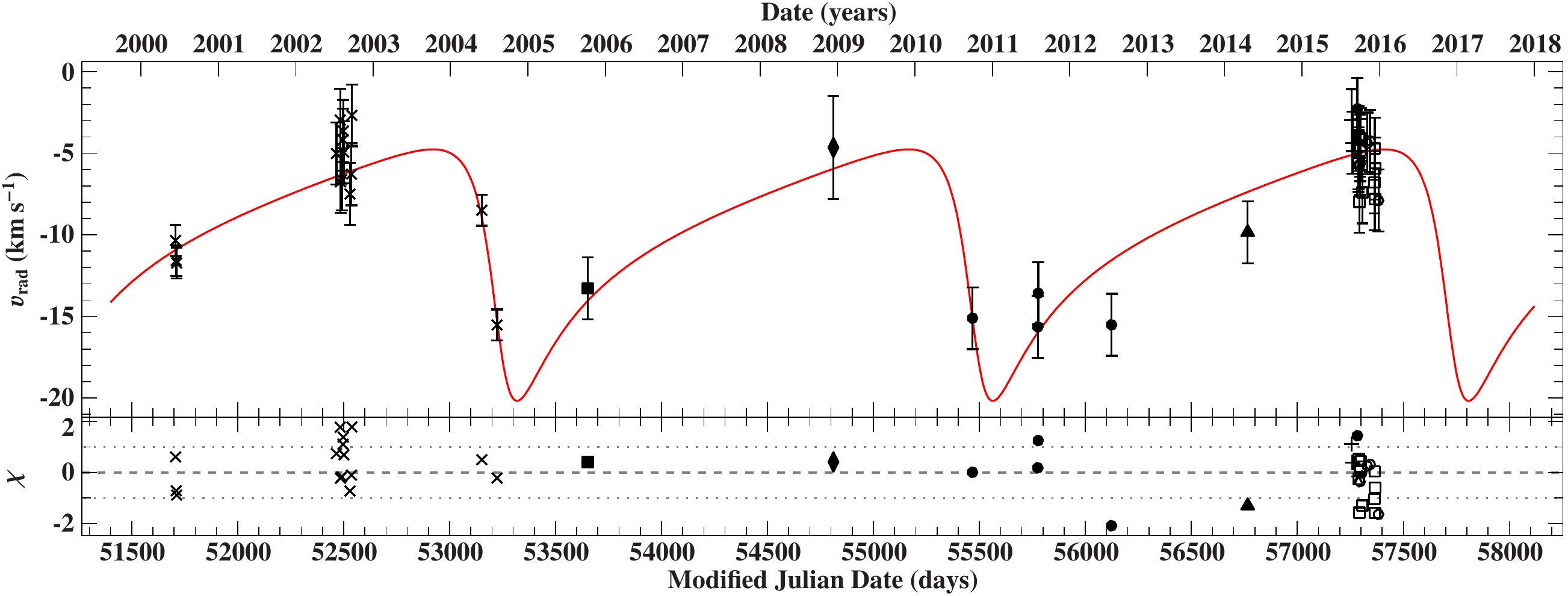}
\caption{Radial velocity curve of 18\,Peg: the measurements are represented by black symbols with error bars while the best-fitting Keplerian model is indicated by the red solid curve. Residuals $\chi$ are shown in the lower panel. The symbols code different instruments, see Table~\ref{table:radial_velocities}.}
\label{fig:vrad_curve}
\end{figure*}
Modeling convective regions in stellar interiors is a long-standing challenge. On the upper main sequence (MS), the hydrogen-burning cores are convective and their sizes are a function of several parameters, in particular the efficiency of convective overshooting. Consequently, the predicted width of the upper MS depends on the numerical treatment of overshooting and the respective choice of a scale length, which can, for example, be calibrated by making use of cluster stars (\citeads{1991A&A...243L...5N}; \citeads{2011A&A...530A.115B}). Nevertheless, large discrepancies in the width of the upper MS still persist between different sets of evolutionary models (compare, e.g., \citeads{2011A&A...530A.115B} and \citeads{2012A&A...537A.146E}). Asteroseismology of pulsating stars has become a powerful tool for probing the internal structure of stars \citepads[e.g.][]{2015AN....336..477A}. Two classes of pulsating stars are found on the upper MS, the $\beta$ Cep stars among the early B-stars (periods of a few hours) and the slowly pulsating B (SPB) stars among late B-types (periods of a few days). As demonstrated by \citetads{2015A&A...580A..27M}, the asteroseismic properties of individual SPB stars can provide important constraints on convective overshooting.

The program star 18\,Peg is a bright ($V=6$\,mag) mid B-type giant (B3\,III) of relatively high Galactic latitude ($l=65.80\degr$, $b=-36.51\degr$). Because of its brightness and early-type spectrum, it is often utilized as a telluric standard or as a background source to investigate the chemical composition of the interstellar medium. In addition, its early-type nature, its unusually low projected rotational velocity $\varv\sin(i_{\mathrm{r}}) = 15 \pm 3$\,km\,s$^{-1}$ \citepads{2012A&A...539A.143N}, which strongly facilitates precise abundance determinations, and its small distance $d = 372 \pm 25$\,pc to the Sun \citepads{2012A&A...539A.143N} make it a prime target to study the present-day chemical abundances in the solar neighborhood, which it seems to be representative of \citepads[see e.g.][]{2012A&A...539A.143N}. It is hence ideally suited for differential abundance analyses to search for chemical peculiarities in, e.g., runaway stars to unravel their origin \citepads{2010ApJ...711..138I}. To our knowledge, there are no remarks in the literature that 18\,Peg might be a pulsator, part of a multiple system, or affected by any other peculiarity. Therefore, it is frequently used as a reference star. Consequently, dozens of unexploited high-resolution ($R = \lambda / \Delta \lambda$), high signal-to-noise (S/N) spectra, which were taken hours, days, weeks, months, and years apart, are available in the archive of the European Southern Observatory (ESO). The high quality of the observations combined with the good time coverage was motivation for us to revisit 18\,Peg for a detailed analysis, which revealed two new facets of the star.

Firstly, the object is actually a single-lined spectroscopic binary (SB1) system. A combined analysis of the derived radial velocity curve and the spectral energy distribution (SED) hints at a wide, eccentric system with a MS or neutron star companion (Sect.~\ref{section:sb1}). Secondly, 18\,Peg shows distortions of its line profiles on a timescale of a few days. These temporal variations are (very likely) caused by slow stellar pulsations (Sect.~\ref{section:spb}). Being one of the most evolved SPB stars currently known, 18\,Peg has the potential to provide a new and rigorous observational constraint on the width of the upper MS (Sect.~\ref{section:benchmark}). Finally, conclusions are summarized (Sect.~\ref{section:summary}).
\section{Single-lined spectroscopic binary}\label{section:sb1}
The search for binaries via spectroscopy is naturally biased towards finding those objects that show large variations of their radial velocity $\varv_{\mathrm{rad}}$ on relatively short timescales, i.e., close systems in which the companion has a significant fraction of the primary's mass. The detection of wide binaries or systems with small orbital inclinations with respect to the observer is more challenging owing to the required long-term monitoring or the small changes in the projected velocity. Consequently, many of these systems remain unrecognized although multiplicity is a common feature particularly among early-type stars \citepads[][and references therein]{2012MNRAS.424.1925C}. It is only because of the high resolution and high S/N ratio of the archival spectra in combination with a sufficient time lag between them that 18\,Peg can be classified without doubt as an SB1 system. 

Based on a detailed analysis of the radial velocity curve (Fig.~\ref{fig:vrad_curve}, see Appendix~\ref{section:rvcurve} for details), we show that 18\,Peg is part of a wide, eccentric binary system with orbital parameters as listed in Table~\ref{table:orbital_params}. The precise nature of the companion, however, remains unknown because there are no direct signatures of the secondary in the optical spectra nor in the SED to pinpoint it. Nevertheless, the available data give rise to the following three conclusions (see Appendix~\ref{section:nature_of_companion} for details): Firstly, the mass function indicates that the companion is more massive than $\sim\! 1\,M_\sun$, hence excluding substellar objects like brown dwarfs or planets. Secondly, the absence of features in the spectra and the SED sets an upper limit on the luminosity of the secondary. Assuming it to be still on the MS and making use of theoretical evolutionary tracks yields a maximum companion mass of $\sim\! 4\,M_\sun$. Thirdly, the invisible component could also be the compact remnant of the binary system's original primary. If so, the derived lifetime of 18\,Peg and statistical considerations about the orbital inclination suggest that the compact object is most likely a neutron star and not a black hole or a white dwarf.
\begin{table}
\renewcommand{\arraystretch}{1.25}
\caption{Orbital parameters.}
\label{table:orbital_params}
\centering
\footnotesize
\begin{tabular}{lr}
\hline\hline
Parameter & Value \\
\hline
Period $P$ & $2245^{+25}_{-30}$\,days \\
Epoch of periastron $T_{\mathrm{periastron}}$ & $57\,730^{+40}_{-60}$\,MJD \\
Eccentricity $e$ & $0.60^{+0.07}_{-0.08}$ \\
Longitude of periastron $\omega$ & $123^{+12}_{-\phantom{0}7}$\,deg \\
Velocity semiamplitude $K_1$ & $7.7^{+1.9}_{-1.1}$\,$\mathrm{km}\,\mathrm{s}^{-1}$ \\
Systemic velocity $\gamma$ & $-9.9\pm0.4$\,$\mathrm{km}\,\mathrm{s}^{-1}$ \\
\hline\hline
Derived parameter & Value \\
\hline
Mass function $f(M)$ & $0.054^{+0.035}_{-0.017}$\,$M_\sun$ \\ 
Projected semimajor axis $a_1 \sin(i)$ & $1.27^{+0.23}_{-0.15}$\,AU \\ 
Projected periastron distance $r_{\mathrm{p}} \sin(i)$ & $108^{+21}_{-17}$\,$R_\sun$ \\ 
\hline
\end{tabular}
\tablefoot{The given uncertainties are single-parameter $1\sigma$-confidence intervals based on $\chi^2$ statistics.}
\end{table}
\begin{table}
\renewcommand{\arraystretch}{1.25}
\caption{Stellar parameters derived from photometry.}
\label{table:photometric_params}
\centering
\footnotesize
\begin{tabular}{lr}
\hline\hline
Photometric parameter & Value \\
\hline
Angular diameter $\Theta$ & $\left(7.13\pm0.05\right)\times10^{-10}$\,rad \\
Color excess $E(B-V)$ & $0.070\pm0.005$\,mag \\
Effective temperature $T_{\mathrm{eff}}$ & $15\,630^{+180}_{-170}$\,K \\
Surface gravity $\log (g\,\mathrm{(cm\,s^{-2})})$ & $3.41^{+0.16}_{-0.14}$\,dex \\
\hline\hline
Derived stellar parameter & Value \\
\hline
Mass $M_1$ & $6.9^{+0.6}_{-0.8}$\,$M_\sun$ \\
Age $\tau$ & $43^{+11}_{-\phantom{0}6}$\,Myr \\
Luminosity $L$ & $4000^{+1300}_{-1400}$\,$L_\sun$ \\
Radius $R_\star$ & $8.6^{+1.4}_{-1.8}$\,$R_\sun$ \\
Distance $d$ & $540^{+\phantom{0}90}_{-110}$\,pc \\
\hline
\end{tabular}
\tablefoot{The given uncertainties are single-parameter $1\sigma$-confidence intervals based on $\chi^2$ statistics. Stellar parameters are derived by comparing the star's position in a $(T_{\mathrm{eff}},\log(g))$ diagram with theoretically predicted evolutionary tracks by \citetads{2012A&A...537A.146E}.}
\end{table}

With respect to the analysis of the SED, we note that the atmospheric parameters deduced here (see Table~\ref{table:photometric_params} and Appendix~\ref{section:nature_of_companion} for details) are in perfect agreement with the calibration by \citetads{1993A&A...268..653N} based on Str\"omgren $uvby\beta$ photometry ($T_{\mathrm{eff}} = 15\,462$\,K, corrected $\log (g) = 3.44$\,dex) and are close to the spectroscopic values ($15\,800 \pm 200$\,K, $3.75 \pm 0.05$\,dex) by \citetads{2012A&A...539A.143N}. Although the spectroscopic analysis of the wings of the Stark-broadened Balmer lines is generally believed to be a better gravity indicator than the line-integrated $H\beta$ index, which is the most sensitive photometric gravity probe, this reveals a tendency to lower $\log (g)$ values that will be important for the discussion in Sect.~\ref{section:benchmark}.
\section{Slowly pulsating B-star}\label{section:spb}\enlargethispage{1\baselineskip}
\begin{figure}
\includegraphics[width=0.49\textwidth]{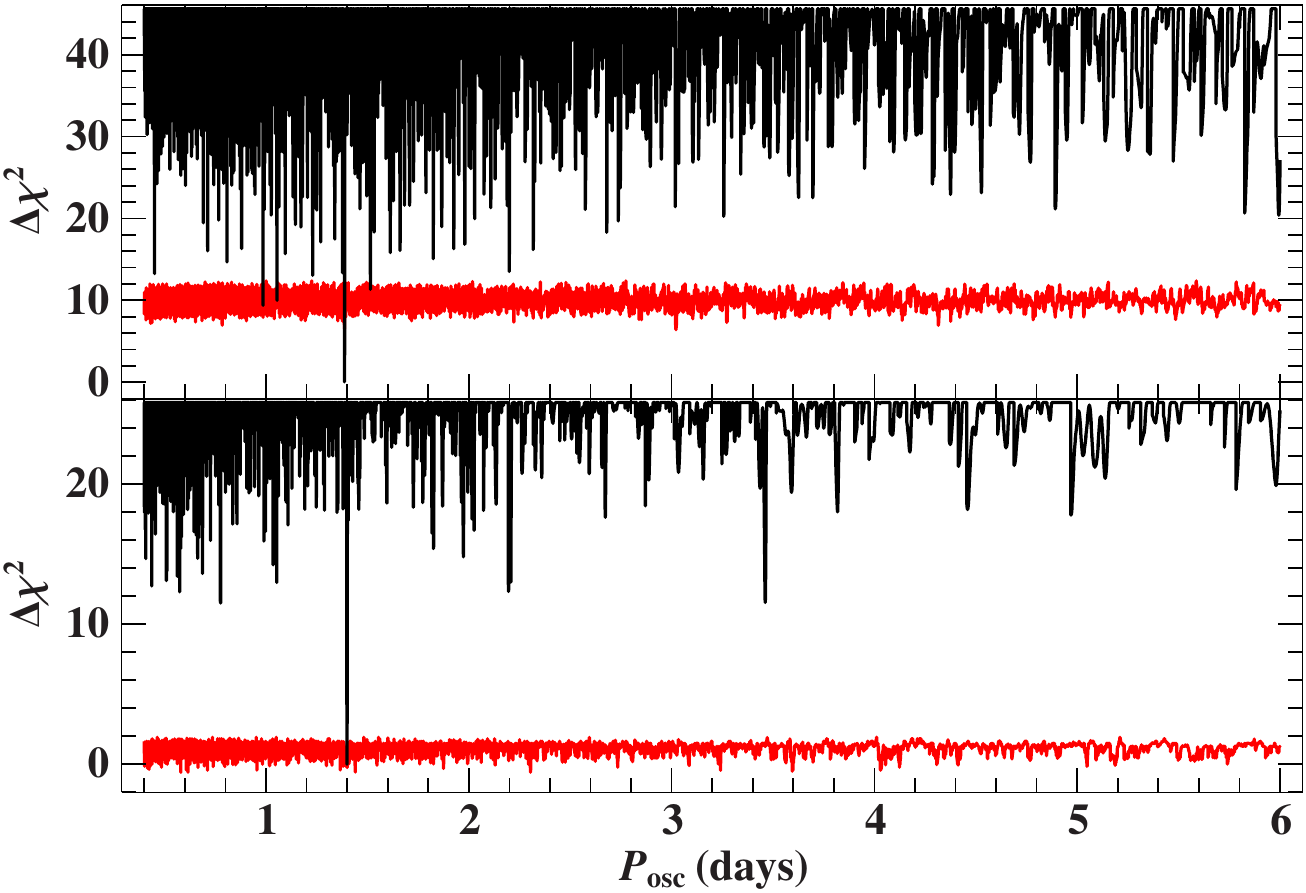}
\caption{The $\chi^2$ landscapes (``periodograms''), which result from fitting the Tycho and {\sc Hipparcos} epoch photometry data (\textit{top}) and the ASAS light-curve (\textit{bottom}) with the model given in Eq.~(\ref{eq:cosine_fit}), as a function of the oscillation period: the differences $\Delta \chi^2$ are with regard to the respective best fit given in Table~\ref{table:oscillation_params}. A step size of $10$ seconds is used to sample the oscillation period. The red line is the $4$\% false-alarm probability threshold deduced from Monte Carlo simulations (see Appendix~\ref{section:photometric_variations} for details).}
\label{fig:periodogram}
\end{figure}

\begin{table}
\caption{\label{table:pulsation_modeling_bestfit}Parameters and derived quantities for the best-fitting pulsational model with $l=5$ and $m=1$.}
\setlength{\tabcolsep}{0.114cm}
\renewcommand{\arraystretch}{1.35}
\centering
\tiny
\begin{tabular}{lr|lr}
\hline\hline
Parameter & Value & Derived quantity & Value \\
\hline
$k^{(0)}$                                      & $0.792^{+0.006}_{-0.007}$                             & $a_{\mathrm{sph}}$               & $0.2688^{+0.0016}_{-0.0009}$\,$R_\sun$ \\
$P_{\mathrm{osc}}$                             & $1.3818\pm0.0001$\,days                       & $\omega^{(0)}$                   & $4.5429\pm0.0002$\,days$^{-1}$ \\
$\phi_{\mathrm{osc,ref}}$                      & $0.4963^{+0.0020}_{-0.0015}$                             & $\Omega/\omega^{(0)}$            & $0.0577^{+0.0010}_{-0.0001}$ \\
$\phi_{\mathrm{rot,ref}}$                      & $0.5323^{+0.0018}_{-0.0020}$                             & $\eta$                           & $0.0042^{+0.0002}_{-0.0001}$ \\
$\Omega/\omega$                                & $0.0576^{+0.0010}_{-0.0001}$                             & $M$                              & $7.3^{+0.2}_{-0.4}$\,$M_\sun$ \\
$\varv\sin(i_{\mathrm{r}})$                    & $16.07^{+0.04}_{-0.03}$\,$\mathrm{km\,s^{-1}}$      & $R_\star$                        & $10.9^{+0.1}_{-0.2}$\,$R_\sun$ \\
${\langle \varv_\mathrm{v}^2 \rangle}{}^{1/2}$ & $1.96^{+0.02}_{-0.01}$\,$\mathrm{km\,s^{-1}}$      & $P_{\mathrm{rot}}$               & $23.9801^{+0.0043}_{-0.3899}$\,days \\
$i_{\mathrm{r}}$                               & $44.2^{+0.2}_{-0.3}$\,$\degr$                    & $\log(g\,\mathrm{(cm\,s^{-2})})$ & $3.22\pm0.01$\,dex \\
\hline
\end{tabular}
\tablefoot{The given uncertainties are single-parameter $1\sigma$-confidence intervals based on a Markov chain Monte Carlo exploration using \texttt{emcee} \citepads{2013PASP..125..306F} and hence only of statistical nature. The meaning of the parameters is explained in Appendix~\ref{section:line_profile_variations}. The surface gravity follows from $g = GMR_\star^{-2}$. The reference epoch for $\phi_{\mathrm{osc,ref}}$ and $\phi_{\mathrm{rot,ref}}$ is $T_{\mathrm{ref}} = 51\,707.23$\,MJD.}
\end{table}

The class of SPB stars was first introduced by \citetads{1991A&A...246..453W} and consists of mid to late B-type stars that show photometric variability on the order of a few days. The pulsations are thought to be driven by an ``opacity bump'' mechanism that excites multi-periodic, non-radial gravity modes with periods in the range $0.4$--$6$\,days and $V$-band amplitudes lower than $0.03$\,mag \citepads[][and references therein]{2015pust.book.....C}. The relatively long periods and low amplitudes make it observationally challenging to find this kind of object. In 2007, the number of confirmed plus candidate Galactic SPB stars was only $116$ \citepads{2007CoAst.150..167D}, clear evidence of their rarity. Empirically, SPB stars tend to be apparently slow rotators --~like 18\,Peg~-- but exceptions do exist \citepads{2007CoAst.150..167D}.

Owing to the extremely high quality of the archival spectra, we detected distortions in the spectral line profiles over a time\-span of a few days (see Fig.~\ref{fig:pulsations}), which closely resemble those of SPB stars. This is consistent with the fact that the position of 18\,Peg in a $(T_{\mathrm{eff}},\log(g))$ diagram lies in the SPB instability domain computed by \citetads{2016MNRAS.455L..67M} when using the spectroscopically derived atmospheric parameters by \citetads{2012A&A...539A.143N}. To examine this further, we independently analyzed Tycho and {\sc Hipparcos} epoch photometry data \citepads{1997ESASP1200.....P} as well as an ASAS \citepads{1997AcA....47..467P} $V$-band light-curve. Both analyses reveal a statistically significant detection of a subtle photometric oscillation with a $V$-band semiamplitude of $\sim\! 0.01$\,mag and a period of $\sim\! 1.4$\,days (see Fig.~\ref{fig:periodogram} and Appendix~\ref{section:photometric_variations} for details). This is in perfect agreement with the properties of SPB stars outlined above. To check whether this oscillation frequency is consistent with the observed changes in the spectral features, the purely dynamical model for pulsationally induced line-profile distortions by \citetads{1997A&AS..121..343S} is fitted to some selected spectra. The outcome of this exercise, which is summarized in Table~\ref{table:pulsation_modeling_bestfit} and shown in Fig.~\ref{fig:pulsation_modeling}, shows that this is indeed the case (see Appendix~\ref{section:line_profile_variations} for details). We note that this preliminary asteroseismic investigation, which is solely based on spectral modeling, already allows us to infer constraints on the stellar mass and radius. Similar to photometry, the deduced value for $\log (g)$ is smaller than the spectroscopic result by \citetads{2012A&A...539A.143N}. Considering the limited database and the simplifications in the model, this finding is only tentative, and continuous spectroscopic and photometric monitoring covering several oscillation periods as well as the use of a more sophisticated asteroseismic model are desirable to confirm or discard it.

By taking both spectroscopic and photometric observations into consideration, it is justified to say that 18\,Peg is most likely a member of the rare class of SPB stars. This conclusion is quite significant because it makes 18\,Peg a prime benchmark star for the width of the upper MS in stellar evolution models.
\begin{figure*}
\centering
\includegraphics[width=1\textwidth]{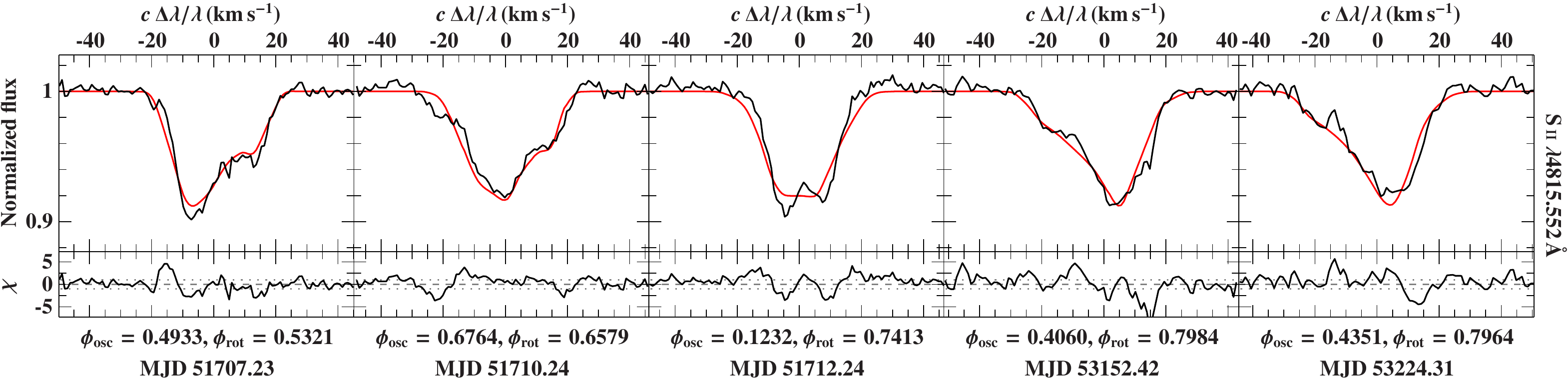}
\caption{Spectral modeling of the pulsationally driven line-profile distortions for five epochs (\textit{columns}) and one exemplary line (see Fig.~\ref{fig:pulsation_modeling_all} for more lines): the {\sc Uves} observations are indicated by a black line, the model (see Appendix~\ref{section:line_profile_variations} for details) by a red one, and the quality of the fit by the residuals $\chi$. Oscillation and rotation phases are listed on the x-axes.}
\label{fig:pulsation_modeling}
\end{figure*}
\section{Potential benchmark star for upper main sequence stellar evolution models}\label{section:benchmark}
\begin{figure}
\centering
\includegraphics[width=0.49\textwidth]{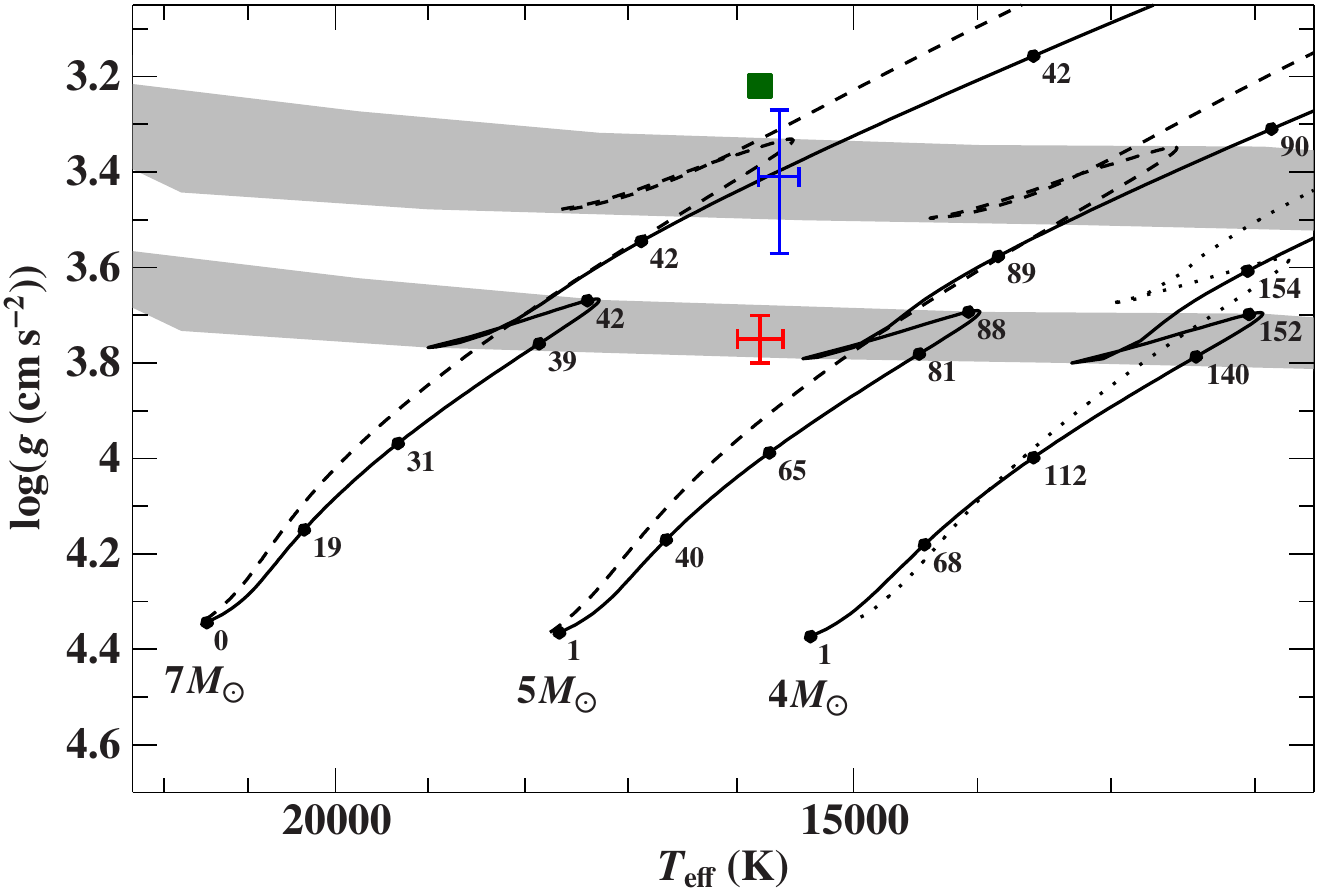}
\caption{Position of 18\,Peg in a $(T_{\mathrm{eff}},\log(g))$ diagram based on spectroscopy \citepads[red error bars;]{2012A&A...539A.143N}, photometry (blue error bars; this work), and preliminary asteroseismology (green square; this work; $T_{\mathrm{eff}}$ taken from spectroscopy). Overlaid are evolutionary tracks for non-rotating stars of solar metallicity and different initial masses by \citetads[][dashed lines]{2011A&A...530A.115B} and \citetads[][solid lines]{2012A&A...537A.146E}. The numbers next to the black filled circles give the evolutionary age in Myr. The gray-shaded areas highlight the transition region between MS and post MS for the two different sets of models. The impact of stellar rotation is demonstrated via the dotted line, which is an \citetads{2012A&A...537A.146E} track for a rotating ($\Omega/\Omega_{\mathrm{crit}}=0.4$) star with an initial mass of $4\,M_\sun$.}
\label{fig:evolution_tracks}
\end{figure}

According to asteroseismology, the terminal-age MS is a hard boundary for the instability strip of SPB stars owing to the very strong damping
of high-order gravity modes in the interiors of post-MS stars \citepads{1999AcA....49..119P}. Under this premise, 18\,Peg is expected to be a MS star given its SPB nature. However, this might be at odds with stellar evolution models that predict a relatively narrow MS band, such as those by \citetads{2012A&A...537A.146E}. While the atmospheric parameters from spectroscopy, which place the star close to the MS turn-off point where a distinction between MS and post MS is impossible owing to the ambiguity of the tracks, could be in line with a MS status, this is certainly not the case for the photometric and asteroseismic analyses (see Fig.~\ref{fig:evolution_tracks}). If our preliminary asteroseismic result is indeed confirmed by better observational data and a more sophisticated model, this would set tight constraints on the width of the upper MS and thus on the efficiency of convective overshooting. We note that rotational mixing, which is the only other mechanism capable of broadening the width of the MS band, is not powerful enough to enlarge the MS by more than $\sim\! 0.1$\,dex (see Fig.~\ref{fig:evolution_tracks}).

By comparing the measured parallax to spectroscopic distance estimates, it is in principle possible to deduce constraints on the surface gravity. Unfortunately, the {\sc Hipparcos} parallax ($3.05 \pm 0.84$\,mas according to \citeads{1997ESASP1200.....P} and $3.04 \pm 0.42$\,mas according to \citeads{2007ASSL..350.....V}) is presumably erroneous since the projected intrinsic binary motion of 18\,Peg within half a year ($\sim\! 0.41$\,AU compared to the motion of Earth of $2$\,AU) was not considered when modeling the parallactic motion. Despite this severe caveat, we note that the {\sc Hipparcos} measurement is best reproduced by the high surface gravity determined via spectroscopy. With the micro-arcsecond accuracy from the \textit{Gaia} satellite \citepads{2001A&A...369..339P}, it will be possible to disentangle binary and parallactic motion and obtain a direct, astrometric constraint on the evolutionary status of 18\,Peg.

Regardless of the precise value of $\log (g)$, 18\,Peg is one of the most evolved SPB stars known to date (compare, e.g., Fig.~1 in \citeads{2007CoAst.150..167D}) and therefore a valuable benchmark object in any case.
\section{Summary}\label{section:summary}
Based on archival data, the standard star 18\,Peg turns out to be a slowly pulsating B-star in a single-lined spectroscopic binary system. Photometric and preliminary asteroseismic analyses indicate that the star has great potential to constrain the width of the upper MS band in stellar evolution models. Spectroscopic and photometric monitoring is required to fully exploit the star's capability as a benchmark object. If observed with the upcoming space mission \textit{TESS} \citepads{2015JATIS...1a4003R}, uninterrupted time-series photometry of high precision for at least a month would be available allowing for a characterization of the pulsational modes and hence for a detailed asteroseismic study.
\begin{acknowledgements}
We thank John E.\ Davis for the development of the {\sc slxfig} module used to prepare the figures in this paper. %
This research has made use of ISIS functions (ISISscripts) provided by 
ECAP/Remeis observatory and MIT (\url{http://www.sternwarte.uni-erlangen.de/isis/}). %
This research used the facilities of the Canadian Astronomy Data Centre operated by the National Research Council of Canada with the support of the Canadian Space Agency. 
Some of the data presented in this paper were obtained from the Mikulski Archive for Space Telescopes (MAST). STScI is operated by the Association of Universities for Research in Astronomy, Inc., under NASA contract NAS5-26555. Support for MAST for non-HST data is provided by the NASA Office of Space Science via grant NNX09AF08G and by other grants and contracts. 
This publication makes use of data products from the Two Micron All Sky Survey, which is a joint project of the University of Massachusetts and the Infrared Processing and Analysis Center/California Institute of Technology, funded by the National Aeronautics and Space Administration and the National Science Foundation. 
This publication makes use of data products from the Wide-field Infrared Survey Explorer, which is a joint project of the University of California, Los Angeles, and the Jet Propulsion Laboratory/California Institute of Technology, funded by the National Aeronautics and Space Administration. 
We thank Markus Schindewolf for his observational efforts and Suhail Masda, Anna Pannicke, and Hartmut Gilbert for their help with the {\sc Flechas} observations at the University Observatory Jena. We are very grateful to Moritz B\"ock, Ulrich Heber, and Norbert Przybilla for valuable discussions in the course of this work and for their suggestions, comments, and contributions to the manuscript.
\end{acknowledgements}
\bibliographystyle{aa}

%
\begin{appendix}
\section{Analysis of the radial velocity curve}\label{section:rvcurve}
\begin{figure*}
\centering
\includegraphics[width=1\textwidth]{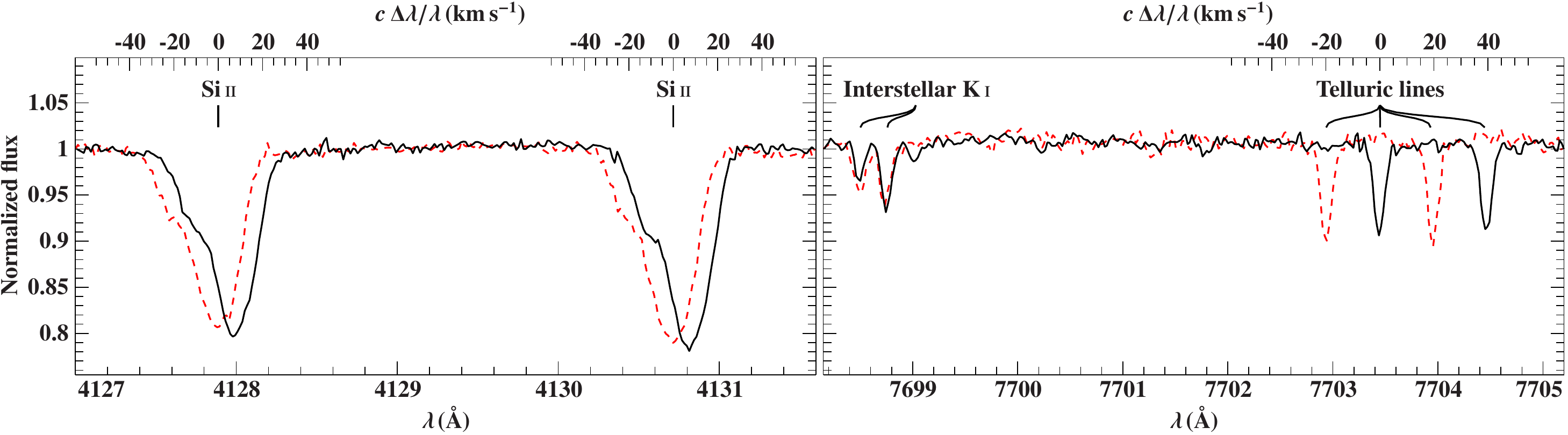}
\caption{Demonstration of a Doppler shift caused by changes in the radial velocity of 18\,Peg: the black solid and red dashed lines are observed {\sc Uves} spectra with $R \approx 55\,000$ taken about $72$\,days apart (MJD $53\,152.42$ and $53\,224.31$, respectively). \textit{Left}: A clear wavelength shift is visible for the stellar Si\,{\sc ii} lines whose asymmetric line profiles are due to slow stellar pulsations (see Sect.~\ref{section:spb}). \textit{Right}: Interstellar K\,{\sc i} and telluric lines are shown to demonstrate that the barycentric correction was properly accounted for.}
\label{fig:velocity_shift}
\end{figure*}
\begin{figure*}
\centering
\includegraphics[width=1\textwidth]{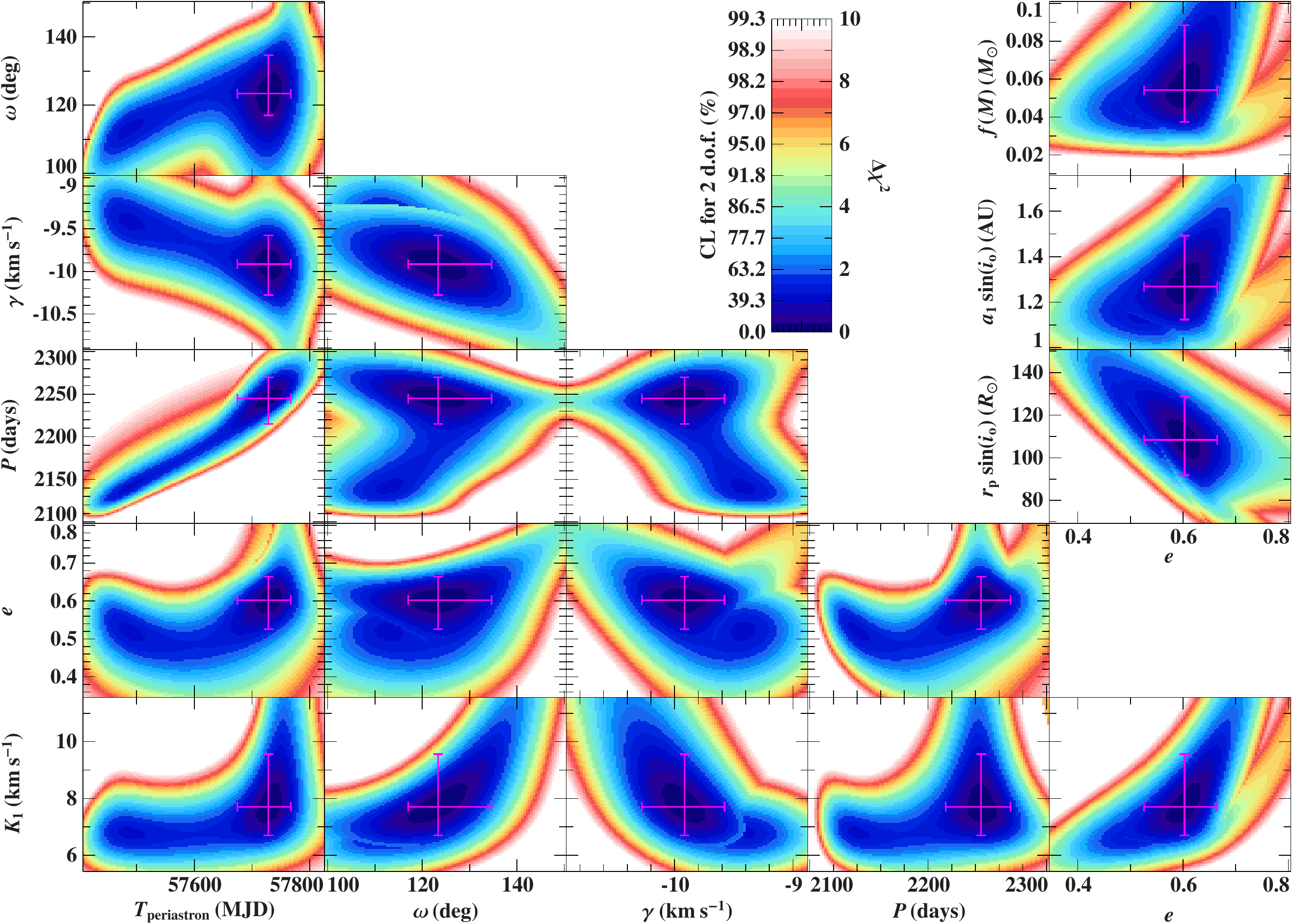}
\caption{Confidence maps for all combinations of orbital parameters plus eccentricity versus mass function, semimajor axis, and periastron distance: the color codes the $\chi^2$ differences $\Delta \chi^2$ with respect to the best fit.  The conversion to joint-confidence levels (CL), which are calculated from the $\Delta \chi^2$ values by means of the cumulative distribution function for 2 degrees of freedom (d.o.f.), is shown as well. The magenta error bars indicate single-parameter $1\sigma$-confidence intervals computed from the condition $\Delta \chi^2 \le 1$.}
\label{fig:vrad_curve_confmaps}
\end{figure*}

Figure~\ref{fig:velocity_shift} illustrates that 18\,Peg is an SB1 system. A long-period double-lined system is excluded, for example by the absence of the three Si\,{\sc ii} lines $\lambda 4128.1$\,{\small \AA}, $\lambda 4130.9$\,{\small \AA}, and $\lambda 6371.4$\,{\small \AA} of the secondary. These spectral features are among the most prominent in the spectra of late B-type or early A-type stars, and their strengths peak at effective temperatures cooler than that of 18\,Peg. Thus, they would be visible in the spectrum if there is a companion that is luminous enough to compete with a B3\,III giant.

To determine the orbital parameters of this system, radial velocities are derived from all {\sc Uves} \citepads{2000SPIE.4008..534D} and X-shooter \citepads{2011A&A...536A.105V} spectra available in the ESO archive\footnote{The 
Reflex workflow \citepads{2013A&A...559A..96F} and the pipeline versions uves-5.5.8 and xshoo-2.6.12 \citepads{2010SPIE.7737E..28M} were used to process those data that are only available in raw format.}. The data so obtained are spread over the years 2000, 2002, 2004, 2010, 2011, 2012, and 2015. Multiple results per night are averaged to reduce the statistical scatter in $\varv_{\mathrm{rad}}$. More data points for the radial velocity curve are extracted from an already available {\sc Foces} \citepads{1998A&AS..130..381P} spectrum (first analyzed by \citeads{2010ApJ...711..138I}), from an ESPaDOnS spectrum \citepads{2006ASPC..358..362D} from the Canadian Astronomy Data Centre\footnote{\url{http://www.cadc-ccda.hia-iha.nrc-cnrc.gc.ca/en/search/}}, and from triggered follow-up observations with {\sc Isis}\footnote{\url{http://www.ing.iac.es/astronomy/instruments/isis/overview/overview.html}} at the William Herschel Telescope, {\sc Twin}\footnote{\url{http://www.caha.es/pedraz/Twin/}} at Calar Alto, {\sc Flechas} \citepads{2014AN....335..417M} at the $90$\,cm telescope of the University Observatory Jena, and {\sc Baches} \citepads{2007Msngr.129...62A} at the $70$\,cm telescope of the public observatory in Dieters\-kirchen\footnote{\url{http://www.sternwarte-dieterskirchen.de/}}. For the high-resolution {\sc Uves} spectra, radial velocities are derived from the Doppler shift of several selected lines, whereas the high-quality {\sc Foces} spectrum ($R = 40\,000$, $\mathrm{S/N}=400$ in the $V$-band, complete spectral coverage between $3860$\,{\small \AA} and $8840$\,{\small \AA}) is taken as a template to determine radial velocities from spectra with lower resolving power by degrading its resolution accordingly and fitting the velocity shift. This technique allows for very precise measurements even from the low-resolution {\sc Isis} ($\Delta \lambda \approx 2${\small \AA}) and {\sc Twin} ($\Delta \lambda \approx 3${\small \AA}) spectra, and from the medium-resolution X-shooter ($R \approx 5000$--$10\,000$), {\sc Flechas} ($R = 9300$, complete spectral coverage between $3900$\,{\small \AA} and $8100$\,{\small \AA}), and {\sc Baches} ($R = 20\,000$, complete spectral coverage between $3960$\,{\small \AA} and $6850$\,{\small \AA}) spectra. Wavelength shifts caused by telescope and instrument flexures in the {\sc Isis} and {\sc Twin} spectra are accounted for by using telluric and interstellar lines as reference points for an absolute wavelength calibration. Finally, there is one measurement with a time stamp in the literature: \citetads{2010ApJ...722..605H} report radial velocities of $-7.8$\,km\,s$^{-1}$ and $-1.5$\,km\,s$^{-1}$ on two successive nights, which gives an average value of $-4.7 \pm 3.2$\,km\,s$^{-1}$.%
\begin{table}
\renewcommand{\arraystretch}{1.07}
\caption{Radial velocity measurements.}
\label{table:radial_velocities}
\centering
\footnotesize
\begin{tabular}{rrrr}
\hline\hline \\[-1em]
MJD & $\varv_{\mathrm{rad}}$ &  Derived phase & Flag \\
(days) & ($\mathrm{km}\,\mathrm{s}^{-1}$) & & \\
\hline
$51\,707.23$ & $-10.3 \pm 1.0$ & $0.3177$ & (1) \\
$51\,710.24$ & $-11.6 \pm 1.0$ & $0.3191$ & (1) \\
$51\,712.24$ & $-11.7 \pm 1.0$ & $0.3200$ & (1) \\
$52\,466.29$ &  $-5.0 \pm 1.9$ & $0.6559$ & (1) \\
$52\,484.27$ &  $-3.0 \pm 1.9$ & $0.6639$ & (1) \\
$52\,485.30$ &  $-6.8 \pm 1.9$ & $0.6643$ & (1) \\
$52\,491.29$ &  $-6.6 \pm 1.9$ & $0.6670$ & (1) \\
$52\,497.17$ &  $-4.2 \pm 1.9$ & $0.6696$ & (1) \\
$52\,498.32$ &  $-3.6 \pm 1.9$ & $0.6701$ & (1) \\
$52\,501.28$ &  $-4.9 \pm 1.9$ & $0.6715$ & (1) \\
$52\,530.09$ &  $-7.5 \pm 1.9$ & $0.6843$ & (1) \\
$52\,537.10$ &  $-6.3 \pm 1.9$ & $0.6874$ & (1) \\
$52\,539.14$ &  $-2.7 \pm 1.9$ & $0.6883$ & (1) \\
$53\,152.42$ &  $-8.5 \pm 1.0$ & $0.9615$ & (1) \\
$53\,224.31$ & $-15.5 \pm 1.0$ & $0.9936$ & (1) \\
$53\,651.88$ & $-13.3 \pm 1.9$ & $0.1840$ & (2) \\
$54\,811.06$ &  $-4.7 \pm 3.2$ & $0.7004$ & (3) \\
$55\,467.16$ & $-15.1 \pm 1.9$ & $0.9927$ & (4) \\
$55\,775.33$ & $-15.7 \pm 1.9$ & $0.1300$ & (4) \\
$55\,777.20$ & $-13.6 \pm 1.9$ & $0.1308$ & (4) \\
$56\,123.18$ & $-15.5 \pm 1.9$ & $0.2849$ & (4) \\
$56\,765.61$ &  $-9.9 \pm 1.9$ & $0.5711$ & (5) \\
$57\,256.10$ &  $-3.0 \pm 1.9$ & $0.7896$ & (6) \\
$57\,259.96$ &  $-4.4 \pm 1.9$ & $0.7913$ & (6) \\
$57\,283.15$ &  $-2.3 \pm 1.9$ & $0.8016$ & (4) \\
$57\,288.81$ &  $-5.3 \pm 1.9$ & $0.8042$ & (7) \\
$57\,289.07$ &  $-5.5 \pm 1.9$ & $0.8043$ & (8) \\
$57\,289.97$ &  $-4.0 \pm 1.9$ & $0.8047$ & (8) \\
$57\,293.01$ &  $-8.0 \pm 1.9$ & $0.8060$ & (8) \\
$57\,294.91$ &  $-5.7 \pm 1.9$ & $0.8069$ & (9) \\
$57\,295.96$ &  $-4.5 \pm 1.9$ & $0.8073$ & (8) \\
$57\,296.93$ &  $-4.8 \pm 1.9$ & $0.8078$ & (8) \\
$57\,297.96$ &  $-4.8 \pm 1.9$ & $0.8082$ & (8) \\
$57\,298.99$ &  $-4.1 \pm 1.9$ & $0.8087$ & (8) \\
$57\,306.94$ &  $-7.4 \pm 1.9$ & $0.8122$ & (8) \\
$57\,329.94$ &  $-4.4 \pm 1.9$ & $0.8225$ & (9) \\
$57\,342.85$ &  $-4.2 \pm 1.9$ & $0.8282$ & (9) \\
$57\,363.77$ &  $-6.8 \pm 1.9$ & $0.8376$ & (8) \\
$57\,364.73$ &  $-4.7 \pm 1.9$ & $0.8380$ & (8) \\
$57\,365.86$ &  $-7.8 \pm 1.9$ & $0.8385$ & (8) \\
$57\,366.73$ &  $-5.9 \pm 1.9$ & $0.8389$ & (8) \\
$57\,382.92$ &  $-7.9 \pm 1.9$ & $0.8461$ & (9) \\
\hline
\end{tabular}
\tablefoot{Uncertainties are $1\sigma$. Data points flagged with (1) are based on {\sc Uves} spectra (and shown as crosses in Fig.~\ref{fig:vrad_curve}), (2) a {\sc Foces} spectrum (filled square), (3) \citetads{2010ApJ...722..605H} (filled diamond), (4) X-shooter spectra (filled circles), (5) an ESPaDOnS spectrum (filled triangle), (6) {\sc Isis} spectra (plus signs), (7) a {\sc Twin} spectrum (multiplication sign), (8) {\sc Flechas} spectra (open squares), (9) {\sc Baches} spectra (open circles).}
\end{table}
\noindent All available $\varv_{\mathrm{rad}}$ data points are listed in Table~\ref{table:radial_velocities} and shown in Fig.~\ref{fig:vrad_curve}. The uncertainties are assumed to be dominated by the distortions of the spectral line profiles due to the slow pulsations (see Sect.~\ref{section:spb}) and thus to be of systematic nature. They are estimated from the condition that the reduced $\chi^2$ at the best fit has to be about $1$. Data points from the five epochs used for the spectral modeling of the pulsations (see Appendix~\ref{section:line_profile_variations}) are more trustworthy and hence assigned an uncertainty that is half as much. The orbital parameters, which were obtained by fitting a Keplerian trajectory to the radial velocity curve, are given in Table~\ref{table:orbital_params}. Additionally, the projected semimajor axis
\begin{equation}
a_1 \sin(i_{\mathrm{o}}) = \frac{1}{2\pi} (1-e^2)^{1/2} K_1 P \,,
\end{equation}
the projected periastron distance
\begin{equation}
r_{\mathrm{p}} \sin(i_{\mathrm{o}}) = (1-e)\, a_1 \sin(i_{\mathrm{o}}) \,,
\end{equation}
and the mass function
\begin{equation}
f(M) \coloneqq \frac{M_2 \sin^3(i_{\mathrm{o}})}{(1+M_1/M_2)^2} = (1-e^2)^{3/2}\frac{K_1^3 P}{2 \pi G} 
\label{eq:mass_function}
\end{equation}
are listed \citepads[see e.g. Eqs.~(2.46) and (2.53) in][]{2001icbs.book.....H}. By parameterizing the Keplerian trajectory in terms of $f(M)$, $a_1 \sin(i_{\mathrm{o}})$, or $r_{\mathrm{p}} \sin(i_{\mathrm{o}})$ instead of, e.g., $K_1$ or $P$, the same standard $\chi^2$ techniques as for the other parameters can be utilized to determine their uncertainties. Figure~\ref{fig:vrad_curve_confmaps} shows confidence maps for all combinations of orbital parameters in order to illustrate possible correlations. The large uncertainties result from the sparse sampling of the orbit that is greater than $6$ years. Further long-term observations are necessary, in particular to better constrain the binary's eccentricity. Nevertheless, it is already obvious from the present data that 18\,Peg is part of a wide, eccentric SB1 system, which is not unusual for early-type stars \citepads[see for instance Fig.~27 in]{2014ApJS..213...34K}.
\section{Nature of the companion}\label{section:nature_of_companion}
There are no features of the secondary component visible in the optical spectra of 18\,Peg. Consequently, all constraints on the nature of the companion are indirect and not very tight. 

\begin{figure}
\centering
\includegraphics[width=0.49\textwidth]{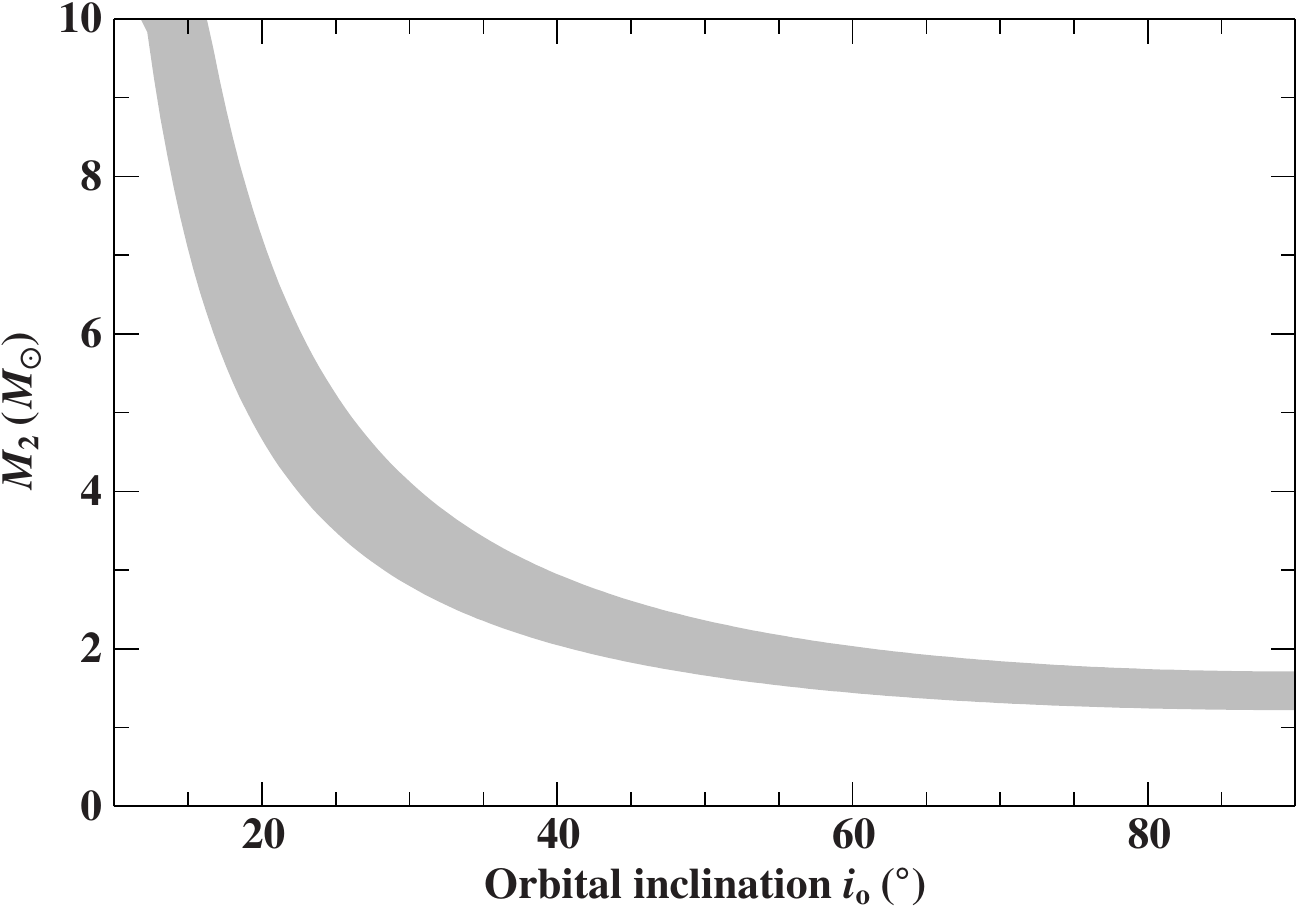}
\caption{Mass of the secondary component as a function of the orbital inclination: a fixed primary mass of $M_1 = 5.8\,M_\sun$ is used to solve the mass function $f(M)$ numerically for $M_2$ (see Eq.~\ref{eq:mass_function}). The width of the shaded region reflects the $1\sigma$-uncertainties of $f(M)$ given in Table~\ref{table:orbital_params}. The solution for $M_2$ is not very sensitive to $M_1$ and the uncertainties of the latter can be neglected here.}
\label{fig:mass_function}
\end{figure}

The mass function $f(M)$, which is derived from the radial velocity curve, is by its definition in Eq.~(\ref{eq:mass_function}) only a lower limit for the secondary mass, $M_2$. However, in the case that the primary mass $M_1$ is known, it can be used to infer a functional relationship between $M_2$ and the orbital inclination $i_{\mathrm{o}}$. This function, which is shown in Fig.~\ref{fig:mass_function} for $M_1 = 5.8\,M_\sun$ \citepads{2014A&A...566A...7N}, is always above $1\,M_\sun$ and thus exceeds the minimum stellar mass of $0.08\,M_\sun$ so that a substellar object like a planet or a brown dwarf can be safely excluded.

\subsection{Main-sequence star}
\begin{table}
\renewcommand{\arraystretch}{1.05}
\caption{Observed and synthetic photometry.}
\label{table:photometry}
\centering
\footnotesize
\begin{tabular}{lrrr}
\hline\hline
Magnitude & Observation & \multicolumn{2}{c}{Model} \\
or color & & Single & Binary \\
\cline{2-4}
& \multicolumn{3}{c}{(mag)} \\
\hline \multicolumn{4}{l}{Johnson-Cousins: (1), (2)} \\
$U-B$ & $-0.568 \pm 0.013$ & $-0.584$ & $-0.580$ \\
$B-V$ & $-0.120 \pm 0.014$ & $-0.099$ & $-0.099$ \\
$V$ & $5.995 \pm 0.008$ & $5.994$ & $5.994$ \\
\hline \multicolumn{4}{l}{Str\"omgren: (3), (4), (5)} \\
$b-y$ & $-0.035 \pm 0.000$ & $-0.008$ & $-0.009$ \\
$m_1$ & $0.081 \pm 0.003$ & $0.058$ & $0.060$ \\
$c_1$ & $0.411 \pm 0.002$ & $0.425$ & $0.427$ \\
$H\beta$ & $2.660 \pm 0.003$ & $2.665$ & $2.671$ \\
\hline \multicolumn{4}{l}{Geneva: (6), (7)} \\
$U-B$ & $0.729 \pm \ldots$ & $0.742$ & $0.745$ \\
$V-B$ & $1.098 \pm \ldots$ & $1.098$ & $1.097$ \\
$B1-B$ & $0.808 \pm \ldots$ & $0.813$ & $0.814$ \\
$B2-B$ & $1.575 \pm \ldots$ & $1.600$ & $1.599$ \\
$V1-B$ & $1.786 \pm \ldots$ & $1.799$ & $1.798$ \\
$G-B$ & $2.306 \pm \ldots$ & $2.319$ & $2.318$ \\
\hline \multicolumn{4}{l}{{\sc Hipparcos: (8), (2)}} \\
$H_p$ & $5.9625 \pm 0.0010$ & $5.9629$ & $5.9625$ \\
\hline \multicolumn{4}{l}{Tycho: (9), (2)} \\
$B_T$ & $5.838 \pm 0.014$ & $5.860$ & $5.860$ \\
$V_T$ & $5.973 \pm 0.010$ & $5.975$ & $5.975$ \\
\hline \multicolumn{4}{l}{2MASS: (10), (11)} \\
$J$ & $6.189 \pm 0.021$ & $6.179$ & $6.180$ \\
$H$ & $6.212 \pm 0.040$ & $6.194$ & $6.195$ \\
$K$ & $6.238 \pm 0.024$ & $6.250$ & $6.251$ \\
\hline \multicolumn{4}{l}{WISE: (12), (13)} \\
$W1$ & $6.282 \pm 0.045$ & $6.276$ & $6.278$ \\
$W2$ & $6.295 \pm 0.022$ & $6.289$ & $6.291$ \\
\hline \multicolumn{4}{l}{Artificial box filter: normalized to Vega} \\
$1300$--$1800$\,{\small \AA} & $4.281 \pm \ldots$ & $4.218$ & $4.217$ \\
$2000$--$2500$\,{\small \AA} & $4.769 \pm \ldots$ & $4.727$ & $4.727$ \\
$2500$--$3000$\,{\small \AA} & $4.758 \pm \ldots$ & $4.754$ & $4.757$ \\
\hline
\end{tabular}
\tablefoot{The quoted uncertainties are taken from literature and are interpreted as $1\sigma$-confidence intervals. Measurements without given uncertainty are assigned the largest of the listed uncertainties of its type, i.e., $0.045$\,mag for magnitudes and $0.014$\,mag for colors. A systematic error of $0.012$\,mag is later added in quadrature to all observations to achieve a reduced $\chi^2$ of about $1$ at the best fit. For each system, the first reference is with respect to the measurements while the following ones are the sources for synthetic photometry, i.e., for the system response functions and the reference magnitudes of Vega, for which the flux-calibrated spectrum alpha\_lyr\_stis\_008.fits from the CALSPEC database (\url{http://www.stsci.edu/hst/observatory/crds/calspec.html}) is used.}
\tablebib{(1)~\citet{mermilliod}; (2)~\citetads{2012PASP..124..140B}; (3)~\citetads{1998A&AS..129..431H}; (4)~\citetads{2011PASP..123.1442B}; (5)~\citetads{2006A&A...454..333C};  (6)~\citetads{1997A&AS..124..349M}; (7)~\citetads{1988A&A...206..357R}; (8)~\citetads{2007ASSL..350.....V}; (9)~\citetads{2000A&A...355L..27H}; (10)~\citetads{2006AJ....131.1163S}; (11)~\citetads{2003AJ....126.1090C}; (12)~\citetads{2012yCat.2311....0C}; (13)~\citetads{2010AJ....140.1868W}.}
\end{table}

The binary is an SB1 system, which allows an upper limit for the luminosity of the companion to be estimated. Consider the line-profile function of a secondary's spectral line (labeled with ``2'') in a composite spectrum $\Phi_{\mathrm{composite,2}}(\lambda)$. It can be expressed by its analogue in a single-star spectrum $\Phi_{\mathrm{single,2}}(\lambda)$ and the continuum luminosities $L_{\mathrm{cont}}(\lambda)$ of the primary (``1'') and secondary component:
\begin{equation}
\Phi_{\mathrm{composite,2}}(\lambda) = \frac{L_{\mathrm{cont,1}}(\lambda)+L_{\mathrm{cont,2}}(\lambda)\,\Phi_{\mathrm{single,2}}(\lambda)}{L_{\mathrm{cont,1}}(\lambda)+L_{\mathrm{cont,2}}(\lambda)} \,.
\end{equation}
Rewriting this in terms of the luminosity ratio gives
\begin{equation}
\frac{L_{\mathrm{cont,2}}(\lambda)}{L_{\mathrm{cont,1}}(\lambda)} = \frac{1-\Phi_{\mathrm{composite,2}}(\lambda)}{\Phi_{\mathrm{composite,2}}(\lambda)-\Phi_{\mathrm{single,2}}(\lambda)} \,.
\label{eq:luminosity_ratio_1}
\end{equation}
Assuming the companion to be an ordinary MS star, one can exploit the three Si\,{\sc ii} lines $\lambda 4128.1$\,{\small \AA} $\lambda 4130.9$\,{\small \AA}, and $\lambda 6371.4$\,{\small \AA} (already mentioned in Appendix~\ref{section:rvcurve}) to evaluate the right-hand side of Eq.~(\ref{eq:luminosity_ratio_1}). The strength of these lines peaks at late B-type or early A-type stars where the maximum central depth in a single-star spectrum --~with high spectral resolution and low rotational broadening~-- is roughly $\Phi_{\mathrm{single,2}}(\lambda_{\rm{center}}) \approx 0.7$. In the ``composite'' spectra shown in Figs.~\ref{fig:velocity_shift} or \ref{fig:pulsations}, the lines of the secondary component are not above the noise level, so $\Phi_{\mathrm{composite,2}}(\lambda_{\rm{center}}) \ge 0.98$ is obtained, which yields
\begin{equation}
\frac{L_2}{L_1} \approx \frac{L_{\mathrm{cont,2}}(\lambda)}{L_{\mathrm{cont,1}}(\lambda)} \approx \frac{L_{\mathrm{cont,2}}(\lambda_{\rm{center}})}{L_{\mathrm{cont,1}}(\lambda_{\rm{center}})} \le \frac{1-0.98}{0.98-0.7} \approx 0.07 \,.
\label{eq:luminosity_ratio_2}
\end{equation}
The choice of the Si\,{\sc ii} lines as proxies for an upper luminosity limit is justified since the luminosity of a star with an effective temperature too low to ionize silicon, i.e., with spectral class later than a mid A-type dwarf, is negligible compared to the B3\,III primary. The ratio derived in Eq.~(\ref{eq:luminosity_ratio_2}) is used later to estimate an upper limit for the mass of the secondary component, provided it is on the MS.

\begin{figure*}
\centering
\includegraphics[width=1\textwidth]{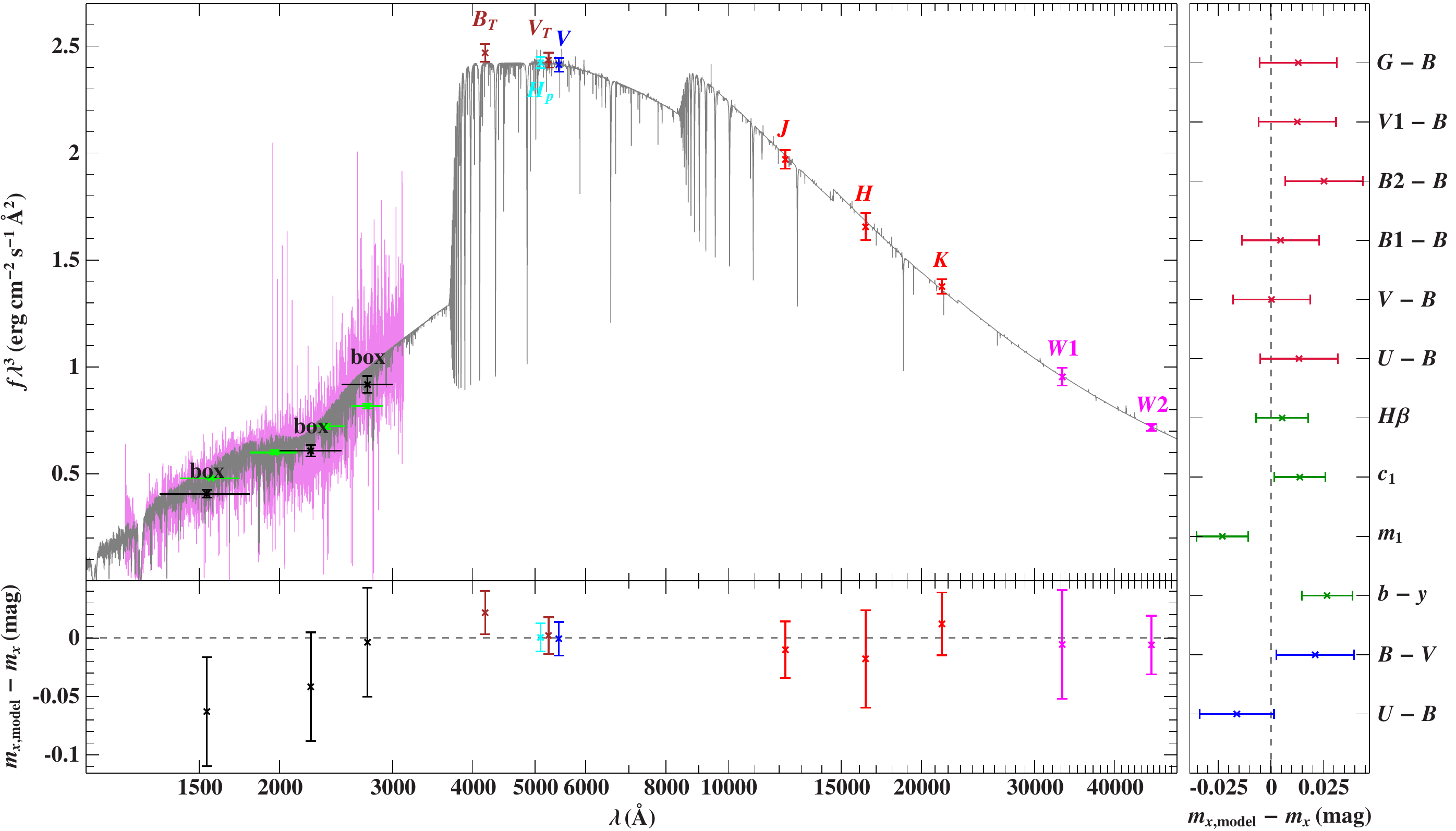}
\caption{Comparison of best-fitting synthetic and observed photometry: The \textit{top panel} shows the spectral energy distribution. The colored data points are fluxes converted from observed magnitudes, while the gray solid line represents the best-fitting single-star model. The three black data points labeled ``box'' are fluxes converted from artificial magnitudes computed by means of box filters of the indicated width from a high-dispersion IUE spectrum (magenta line). The four lime green ultraviolet data points are flux measurements from the TD1 catalog \citepads{1978csuf.book.....T} and are shown here for a consistency check. The residual panels at the \textit{bottom}/\textit{side} show the differences between synthetic and observed magnitudes/colors. The photometric systems have the following color code: Johnson-Cousins (blue), Str\"omgren (green), Tycho (brown), {\sc Hipparcos} (cyan), 2MASS (red), WISE (magenta), Geneva (crimson).}
\label{fig:photometry}
\end{figure*}

\begin{figure}
\centering
\includegraphics[width=0.49\textwidth]{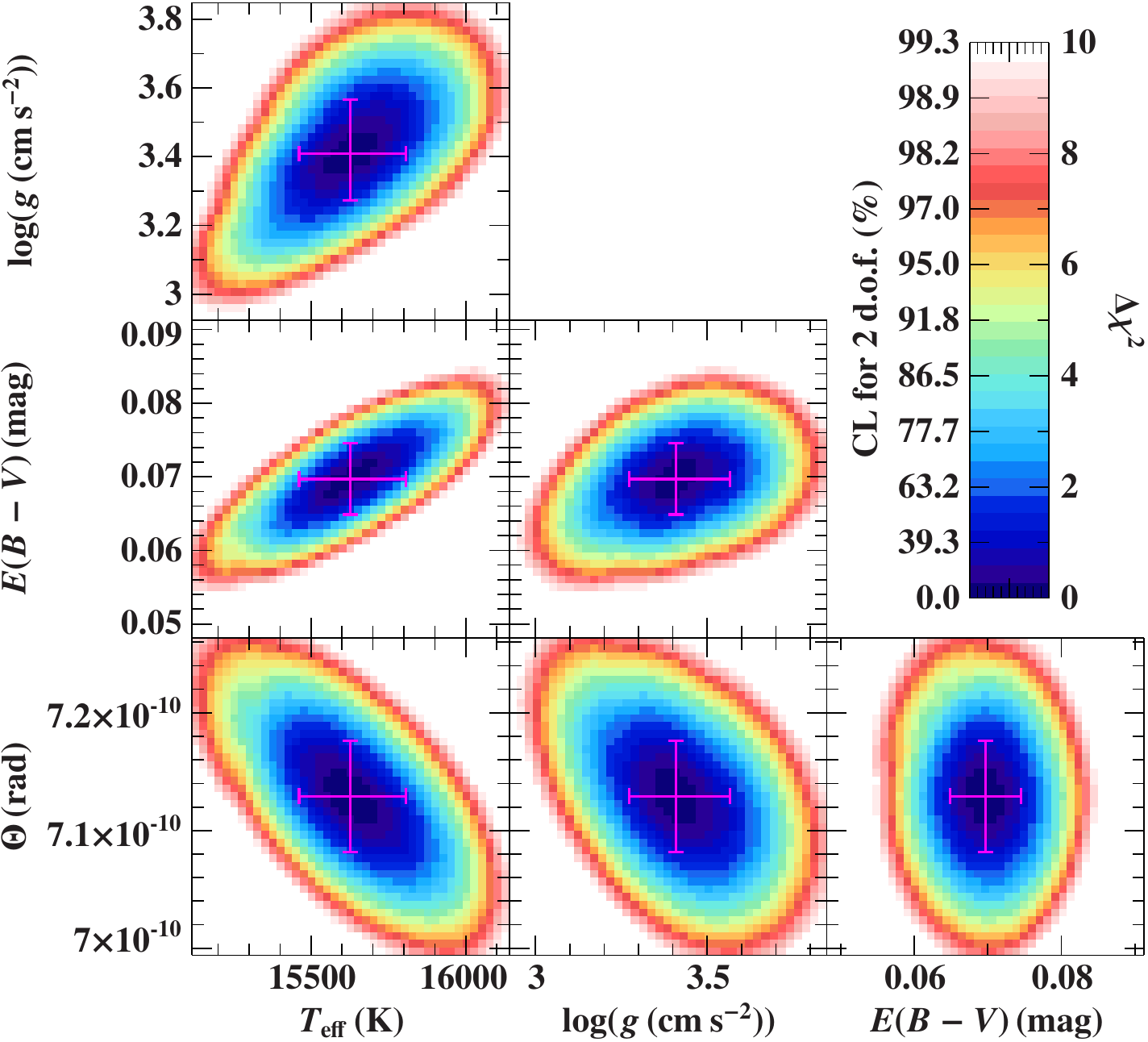}
\caption{Confidence maps for all combinations of photometric parameters: the meaning of the color code and the magenta error bars is the same as in Fig.~\ref{fig:vrad_curve_confmaps}.}
\label{fig:photometry_confmaps}
\end{figure}

The SED, which can be very useful to find cool companions by virtue of an infrared excess, is also studied here. To do so, 21 photometric data points covering the visual and infrared regime are compiled from literature. Additionally, three artificial magnitudes in the ultraviolet are computed via box filters from a high-dispersion spectrum taken with the International Ultraviolet Explorer (IUE)\footnote{Unfortunately, the quality of the available IUE spectra is not sufficient to measure Doppler shifts at the level of accuracy and precision required to improve the radial velocity curve in Fig.~\ref{fig:vrad_curve}.} and publicly available in the MAST archive\footnote{\url{http://archive.stsci.edu/}} under data IDs SWP20593 and LWR16508. Table~\ref{table:photometry} lists the observations and the references for synthetic photometry (system response functions, reference magnitudes) for all passbands employed here. Synthetic SEDs as a function of effective temperature and surface gravity are interpolated from the TLUSTY BSTAR2006 grid \citepads{2007ApJS..169...83L} with solar chemical composition and a microturbulence of $2$\,km\,s$^{-1}$. The decrease in flux with increasing distance $d$ is parametrized via the angular diameter $\theta$ of the stellar disk with radius $R_\star$, $\theta = 2 R_\star / d$. Interstellar extinction is accounted for by applying a reddening correction $A(\lambda)$, which is the extinction in magnitude at wavelength $\lambda$. \citetads{1999PASP..111...63F} provides expressions for $A(\lambda)$ as a function of the color excess $E(B-V)$, which is a measure for the quantity of absorbers, and the extinction parameter $R_V = A(V) / E(B-V)$, which is a probe for the properties of the absorbing matter. For weakly reddened objects like 18\,Peg, it is impossible to distinguish between the two parameters making it appropriate to assume $R_V = 3.1$ \citepads[the standard value for the diffuse interstellar medium, see for example][]{1999PASP..111...63F}. The results obtained from fitting a single-star SED to the data points are summarized in Table~\ref{table:photometric_params} and shown in Fig.~\ref{fig:photometry}. Confidence maps are presented in Fig.~\ref{fig:photometry_confmaps}. With no signatures of a companion showing up in the residual panels of Fig.~\ref{fig:photometry}, the observed SED is nicely reproduced by a single-star model and, thus, of little help in unraveling the nature of the secondary component. However, by assuming that the latter is a MS star of the same age as 18\,Peg and by exploiting the luminosity limit determined via spectroscopy, it is possible to have some clues. To this end, isochrones \citepads{2012A&A...537A.146E}, starting at the 18\,Peg position in the $(T_{\mathrm{eff}},\log(g))$ diagram, are followed towards lower masses until Eq.~(\ref{eq:luminosity_ratio_2}) is fulfilled. The resulting parameter limits are listed in Table~\ref{table:limits_secondary}. To check whether such a companion is compatible with photometry, the magnitudes and colors of a binary spectrum composed of a primary and a secondary component with parameters shown in Table~\ref{table:photometric_params} and \ref{table:limits_secondary}, respectively, are computed --~after re-fitting $\Theta$ and $E(B-V)$~-- and are compared to the single-star model in Table~\ref{table:photometry}. The differences are negligible, which shows that the derived limits in Table~\ref{table:limits_secondary} are consistent with the observed SED. In conclusion, a MS star with a mass of $1$--$4\,M_\sun$ fulfills all the constraints set by spectroscopy, photometry, and the radial velocity curve and is therefore a promising candidate for the unseen companion.

\begin{table}
\renewcommand{\arraystretch}{1.05}
\caption{Derived parameter limits for a main-sequence companion.}
\label{table:limits_secondary}
\centering
\footnotesize
\begin{tabular}{lrr}
\hline\hline
Parameter & & Value \\
\hline
Mass $M_2$ & $\le$ & $4.06$\,$M_\sun$ \\
Effective temperature $T_{\mathrm{eff},2}$ & $\le$ & $14\,870$\,K \\
Surface gravity $\log (g_2\,\mathrm{(cm\,s^{-2})})$ & $\ge$ & $4.22$\,dex \\
Luminosity $L_2$ & $\le$ & $277$\,$L_\sun$ \\
Radius $R_{\star,2}$ & $\le$ & $2.58$\,$R_\sun$ \\
\hline
\end{tabular}
\end{table}
\subsection{Compact object}
Alternatively, the secondary component could also be the stellar remnant of the system's original primary, i.e., a star more massive and thus more short-lived than 18\,Peg. Depending on the initial mass of that hypothetical star, the companion could be either a neutron star or a black hole. A white dwarf is rather unlikely because, on the one hand, stellar evolution predicts that the maximum progenitor mass of such an object is $\sim\! 8\,M_\sun$. On the other hand, according to the evolutionary tracks by \citetads{2012A&A...537A.146E}, the minimum mass of a star with a lifetime shorter than the current age of 18\,Peg is just slightly below $8\,M_\sun$ leaving only a tiny mass window for the white dwarf scenario. The presence of a neutron star or a black hole would imply that the binary system has survived a supernova explosion. 

Owing to the low or nonexistent intrinsic luminosities of neutron stars or black holes, there is no chance of seeing these objects next to a B3 giant --~unless there is mass accretion onto the compact object that leads to the emission of X-rays. For significant mass transfer to occur, the distance between the donor and accretor has to be quite small. Because it is a very wide binary system with a non-extreme orbital eccentricity, this condition is not met by 18\,Peg for which periastron passages are probably larger than $90\,R_\sun$ (see $e$ versus $r_{\mathrm{p}} \sin(i_{\mathrm{o}})$ in Fig.~\ref{fig:vrad_curve_confmaps}).

All currently available observations are consistent with a MS and with a compact companion. Only if the orbital inclination $i_{\mathrm{o}}$ of the system is smaller than $\sim\! 20\degr$ could a MS star be excluded: According to Fig.~\ref{fig:mass_function}, the secondary mass would then be larger than $\sim\! 4\,M_\sun$, which is the estimated upper mass limit for a MS secondary (see Table~\ref{table:limits_secondary}). Since this number exceeds the Tolman-Oppenheimer-Volkoff limit, which is the maximum mass of a neutron star, a black hole would be the only remaining option in that case. Unfortunately, it is impossible to directly constrain $i_{\mathrm{o}}$ except for those systems that exhibit eclipses. Assuming an isotropic distribution and thus a spherically averaged value for $\sin^3(i_{\mathrm{o}})$ of $3\pi/16$ as well as a primary mass of $5.8\,M_\sun$, the most likely secondary mass is $1.73^{+0.39}_{-0.24}\,M_\sun$. This is larger than the Chandrasekhar mass limit of $1.46\,M_\sun$ and is close to the mass of most neutron stars known to date \citepads{2013ApJ...778...66K}. Statistically speaking, the companion is hence not massive enough to be a black hole and is probably either a MS or a neutron star.

\section{Photometric variations}\label{section:photometric_variations}
\begin{figure}
\includegraphics[width=0.49\textwidth]{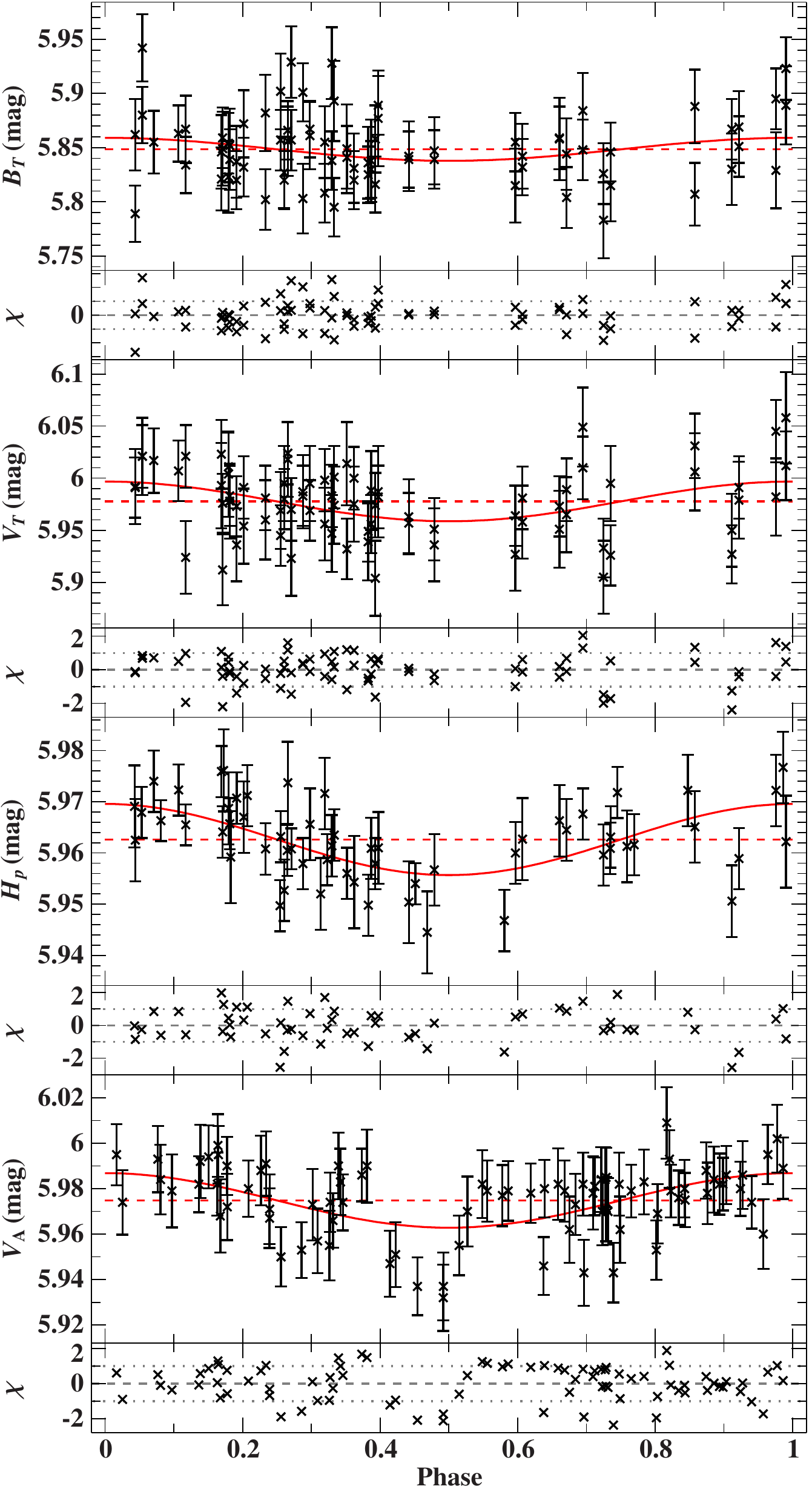}
\caption{Phased light-curves of Tycho and {\sc Hipparcos} epoch photometry and ASAS data: the measurements are represented by black crosses with error bars while the best-fitting model (see Table~\ref{table:oscillation_params}) is indicated by the red solid curve. The red dashed line marks the derived mean magnitude. Residuals $\chi$ are shown as well.}
\label{fig:phased_lightcurves}
\end{figure}

\begin{table}
\renewcommand{\arraystretch}{1.05}
\caption{Oscillation parameters.}
\label{table:oscillation_params}
\centering
\footnotesize
\begin{tabular}{lr}
\hline\hline
Parameter & Value \\
\hline
\multicolumn{2}{l}{Tycho and {\sc Hipparcos} epoch photometry data:} \\
Period $P_{\mathrm{osc}}$ & $1.38711\pm0.00014$\,days \\
Reference epoch $T_{\mathrm{ref}}$ (fixed) & $47\,898.49$\,MJD \\
Phase $\phi_{\mathrm{osc,ref}}$ at epoch $T_{\mathrm{ref}}$ & $0.58\pm0.05$ \\
$V_T$ mean magnitude & $5.848\pm0.004$\,mag \\
$V_T$ semiamplitude & $0.011\pm0.006$\,mag \\
$B_T$ mean magnitude & $5.978\pm0.004$\,mag \\
$B_T$ semiamplitude & $0.019\pm0.006$\,mag \\
$H_p$ mean magnitude & $5.9626\pm0.0009$\,mag \\
$H_p$ semiamplitude & $0.0069\pm0.0013$\,mag \\
\hline
\multicolumn{2}{l}{ASAS light-curve:} \\
Period $P_{\mathrm{osc}}$ & $1.39976\pm0.00030$\,days \\
Reference epoch $T_{\mathrm{ref}}$ (fixed) & $54\,229.40$\,MJD \\
Phase $\phi_{\mathrm{osc,ref}}$ at epoch $T_{\mathrm{ref}}$ & $0.68\pm0.05$ \\
$V_{\mathrm{A}}$ mean magnitude & $5.9748\pm0.0016$\,mag \\
$V_{\mathrm{A}}$ semiamplitude & $0.0120\pm0.0024$\,mag \\
\hline
\end{tabular}
\tablefoot{The given uncertainties are single-parameter $1\sigma$-confidence intervals based on $\chi^2$ statistics.}
\end{table}
To check whether there are pulsational signatures in the light curve of 18\,Peg, two different datasets are analyzed. The first consists of Tycho and {\sc Hipparcos} epoch photometry data \citepads{1997ESASP1200.....P}. With $59$ photometric measurements in the {\sc Hipparcos} $H_p$-band and $80$ data points in each of the two Tycho bands ($B_T$, $V_T$), all of which are spread over roughly $1000$\,days, the sampling is quite sparse. The same applies to the second dataset, which is an ASAS \citepads{1997AcA....47..467P} $V$-band ($V_{\mathrm{A}}$) light-curve with $85$ observations in $569$\,days. Given the brightness of 18\,Peg, the magnitudes derived from the largest aperture ($6$ pixels in diameter) are employed here. Data points from ASAS not flagged with ``A'' (=best data) or taken before MJD\,$54\,229$ are omitted because of several obvious outliers. To identify a single pulsational mode, a model curve of the simplest form
\begin{equation}
\mathrm{mag}_j(t) = \overline{\mathrm{mag}}_j + A_j \cos\left(2\pi\left[(t-T_{\mathrm{ref}})/P_{\mathrm{osc}}+\phi_{\mathrm{osc,ref}}\right] \right)
\label{eq:cosine_fit}
\end{equation}
with a time-dependent magnitude $\mathrm{mag}_j(t)$, a mean magnitude $\overline{\mathrm{mag}}_j$, and an oscillation semiamplitude $A_j$ is chosen. The parameter $\phi_{\mathrm{osc,ref}}$ is the phase at the fixed reference epoch $T_{\mathrm{ref}}$. The index $j \in \{V_T, B_T, H_p, V_{\mathrm{A}}\}$ refers to the four available passbands. The two datasets are independently fitted using $\chi^2$ minimization tools provided by the Interactive Spectral Interpretation System \citepads{2000ASPC..216..591H}. To identify the dominant mode, the oscillation period $P_{\mathrm{osc}}$ is sampled from $0.4$--$6$\,days in steps of $10$ seconds while all other parameters are fitted. The resulting periodograms are shown in Fig.~\ref{fig:periodogram} and reveal an outstanding detection at $P_{\mathrm{osc}} \sim\! 1.4$\,days. The corresponding best-fitting model parameters are listed in Table~\ref{table:oscillation_params} and the respective phased light-curves are shown in Fig.~\ref{fig:phased_lightcurves}. We note that the reduced $\chi^2$ at the best fit of the Tycho and {\sc Hipparcos} data is $1.005$. It is therefore not necessary to make the function in Eq.~(\ref{eq:cosine_fit}) more complex, for example by including more oscillation terms, in order to properly model the data. The given uncertainties for the ASAS data are not individual photometric errors but averaged ones for the respective frame. Their mean value ($0.037$\,mag) exceeds the standard deviation of the actual measurements ($0.016$\,mag), which indicates that the stated numbers are overestimated. The original uncertainties are thus rescaled by a factor of $0.140^{1/2}$ to give a reduced $\chi^2$ at the best fit of $1$ instead of $0.140$.

As expected for SPB stars, the resulting oscillation semiamplitudes are very small. That is why it is important to check whether the detection is statistically significant or just a coincidence caused by noise fluctuations in a constant underlying signal. To this end, Monte Carlo simulations are performed. A total of $1000$ mock light-curves are created for each of the four passbands by replacing the measured magnitudes by Gaussian random numbers with the magnitudes' mean value as center and the magnitudes' standard deviation as width. The mock data are then analyzed as described before. In particular, $1000$ periodograms like those shown in Fig.~\ref{fig:periodogram} are computed for both datasets. For each individual period, the distribution of $\Delta \chi^2$ values --~which are still with regard to the best fit given in Table~\ref{table:oscillation_params}~-- is constructed from the sample of mock periodograms. From each of these distributions, the $4$\%-quantile ${\Delta \chi^2}_{4\%}(P_{\mathrm{osc}})$ is determined. The probability of obtaining a $\Delta \chi^2$ value that is smaller than this quantity is thus less than $4$\% if the light curve is indeed a constant one with noise. Consequently, ${\Delta \chi^2}_{4\%}(P_{\mathrm{osc}})$ is the $4$\% false-alarm probability threshold. Given its location in Fig.~\ref{fig:periodogram}, which is indicated by the red line, we can conclude that the detection at $P_{\mathrm{osc}} \sim\! 1.4$\,days is significant with more than $96$\% confidence.
\section{Line-profile variations}\label{section:line_profile_variations}
\begin{figure*}
\centering
\includegraphics[width=1\textwidth]{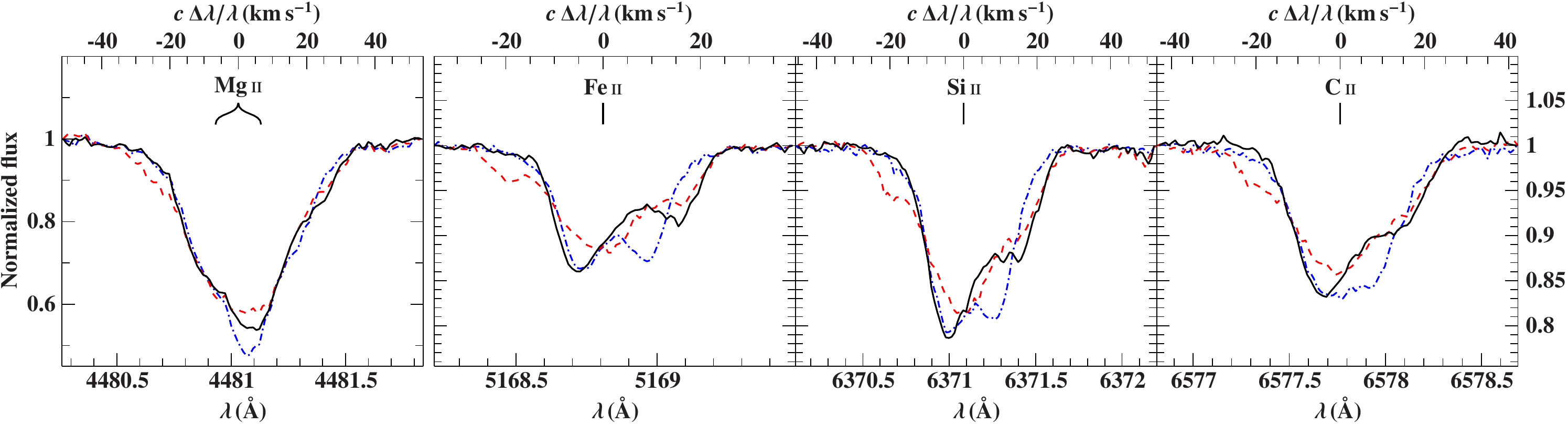}
\caption{Demonstration of line-profile variations by means of prominent metal lines: the black solid, red dashed, and blue dash-dotted lines are observed {\sc Uves} spectra with spectral resolutions $R = 77\,810$ (Mg\,{\sc ii}) and $R = 107\,200$ (Fe\,{\sc ii}, Si\,{\sc ii}, and C\,{\sc ii}) taken three and two days apart (MJD $51\,707.23$, $51\,710.24$, and $51\,712.24$). All lines --~we note that Mg\,{\sc ii} is a blend of two lines of similar strength~-- show the same temporal and morphologic changes, which resemble those of SPB stars.}
\label{fig:pulsations}
\end{figure*}

The detection of distortions in the spectral line profiles over a timespan of a few days (see Fig.~\ref{fig:pulsations}) hints at an SPB nature of 18\,Peg. In principle, a short-period double-lined spectroscopic binary (SB2) system with two components of similar luminosity could also be responsible for these variations. However, the change in the line profile from the black to the blue curve in Fig.~\ref{fig:pulsations} is barely compatible with the expected reflex motion in an SB2 system since only the bump at longer wavelengths moved, whereas its counterpart at shorter wavelengths kept still. Additionally, the expected velocity semiamplitudes in such a system are much larger than those indicated by the small Doppler shifts in Fig.~\ref{fig:pulsations}. For instance, the orbital velocity in a system composed of two $5.8\,M_\sun$ stars, which is the mass of 18\,Peg according to \citetads{2014A&A...566A...7N}, on a circular orbit with a $3$-day period is $167$\,km\,s$^{-1}$. To reconcile this with the tiny displacements observed in Fig.~\ref{fig:pulsations}, an orbital inclination $i_{\mathrm{o}}$ of less than $\sim\! 7\degr$ would be required, i.e., the system had to be seen almost pole-on, which is possible but unlikely from a statistical point of view. A similar reasoning applies to moving spots on the stellar surface, which can also lead to time-dependent distortions of spectral line profiles (see, e.g., \citeads{2001A&A...380..177B}, \citeyearads{2004A&A...413..273B}). Based on a stellar radius of $R_\star = 5.5\,R_\sun$ \citepads{2014A&A...566A...7N}, an upper limit for the rotation period of the star is given by $P_{\mathrm{rot}}/\sin(i_{\mathrm{r}}) = 2 \pi R_\star / \varv\sin(i_{\mathrm{r}}) \approx 18.6$\,days, which would imply a rotational inclination $i_{\mathrm{r}}$ of less than $\sim\! 9\degr$ to obtain spot-induced spectral variations with a period of $<3$\,days. For such low values of $i_{\mathrm{r}}$, however, the temporal effects of stellar spots on the spectrum are expected to vanish because an observer on Earth will then always see the same hemisphere and thus a constant radiation of the star. Stellar pulsations yield a simpler explanation for the spectral variations.

\begin{figure*}
\centering
\includegraphics[width=1\textwidth]{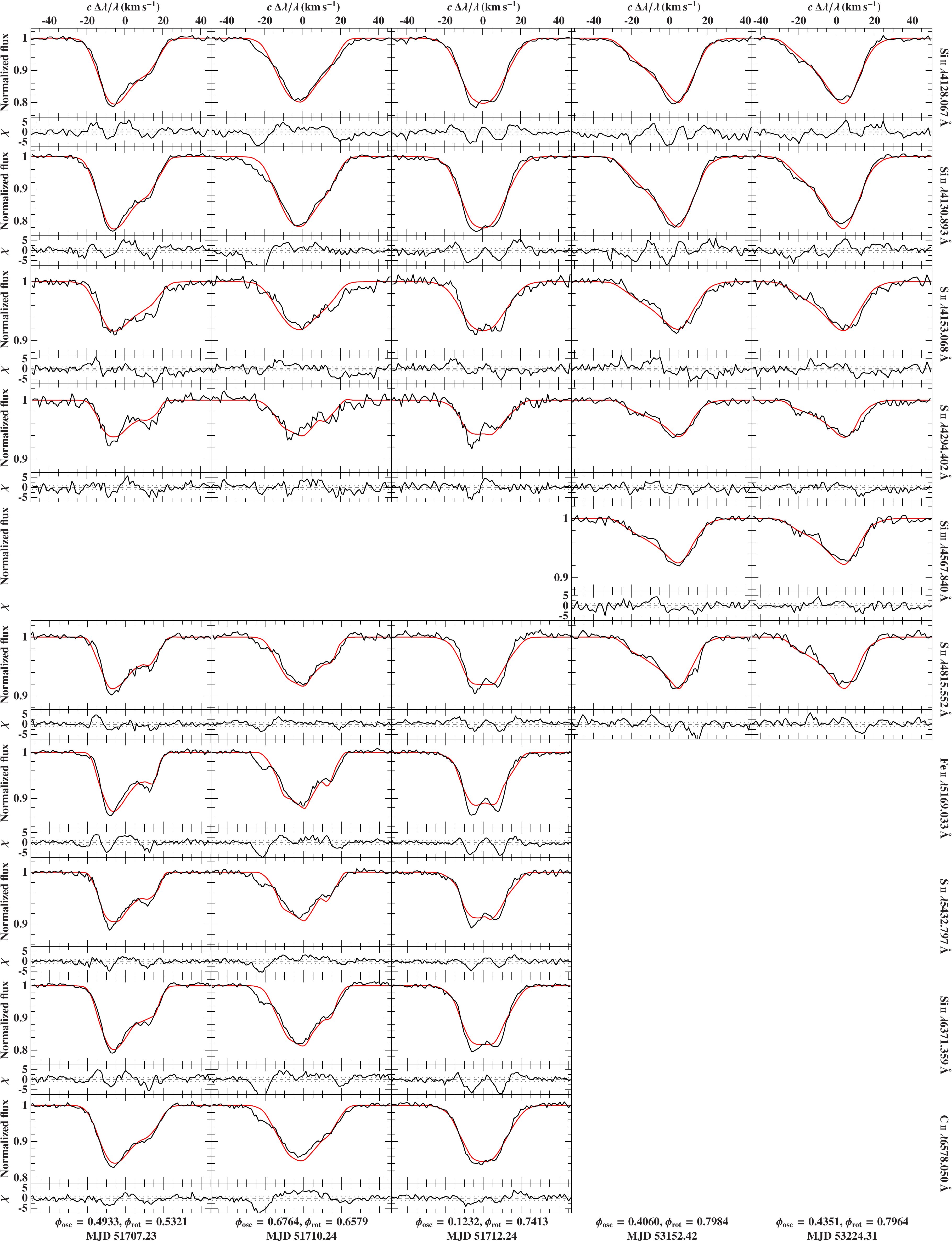}
\caption{Same as Fig.~\ref{fig:pulsation_modeling}, but for all ten metal lines (\textit{rows}). Empty slots result from varying instrumental setups.}
\label{fig:pulsation_modeling_all}
\end{figure*}

\citetads{1997A&AS..121..343S} provide a formulation of the surface-velocity field of a rotating, adiabatically pulsating star whose pulsational and rotational axes are aligned. By numerically integrating the respective Doppler shifts of this field, it is possible to synthesize spectral line profiles that account for the pulsational motion. We note that atmospheric variations such as changes in the temperature are neglected in this purely dynamical approach.  Following \citetads{1997A&AS..121..343S}, the resulting pulsational Doppler profile $\Phi_{\mathrm{osc}}$ is a function of the degree $l$ and azimuthal order $m$ of the oscillation mode, the vertical amplitude $a_{\mathrm{sph}}$, the ratio of the horizontal and vertical amplitude $k^{(0)}$ (superscripts $^{(0)}$ refer to quantities in the non-rotating case), the angular oscillation frequency $\omega^{(0)}$, the ratio of the angular rotation frequency $\Omega$ and $\omega^{(0)}$, the inclination of the pulsational/rotational axis $i_{\mathrm{r}}$, the projected rotational velocity $\varv\sin(i_{\mathrm{r}})$, the oscillation phase $\phi_{\mathrm{osc}}$, and the rotation phase $\phi_{\mathrm{rot}}$:
\begin{equation}
\Phi_{\mathrm{osc}} = \Phi_{\mathrm{osc}}(l,m,a_{\mathrm{sph}},k^{(0)},\omega^{(0)}, \Omega/\omega^{(0)}, i_{\mathrm{r}},\varv\sin(i_{\mathrm{r}}),\phi_{\mathrm{osc}},\phi_{\mathrm{rot}})\,.
\label{eq:pulsational_profile}
\end{equation}
For convenience, the parameters $\omega^{(0)}$ and $\Omega/\omega^{(0)}$ are replaced here by $P_{\mathrm{osc}} = 2\pi/\omega$ and $\Omega/\omega$ exploiting Eq.~(11) in \citetads{1997A&AS..121..343S}. Moreover, $a_{\mathrm{sph}}$ is substituted with the square root of the mean square of the vertical velocity component when averaged over a spherical surface and oscillation period ${\langle \varv_\mathrm{v}^2 \rangle}{}^{1/2} = \omega\,a_{\mathrm{sph}}(8\pi)^{-1/2}$ (compare Eq.~(3.136) in \citeads{2010aste.book.....A}). The distortions of the spectral lines due to pulsation affect primarily narrow metal lines, whose individual profile (labeled with index $j$) can be well approximated by a Gaussian function with area $A_j$, width $W_j$, and central wavelength $C_j$:
\begin{equation}
\Phi_{j\mathrm{,Gauss}}(A_j,W_j,C_j) = 1 - \frac{A_j}{\sqrt{2\pi} W_j} \exp\left(-\frac{1}{2}\left(\frac{\lambda-C_j}{W_j}\right)^2\right) \,.
\label{eq:gauss_profile}
\end{equation}
The convolution of this Gaussian curve with the pulsational Doppler profile gives the final model $\Phi_{j\mathrm{,model}}$, which can be fitted to an observed metal absorption line. The only time-dependent quantities entering $\Phi_{j\mathrm{,model}}$ are 
$\phi_{\mathrm{osc}}(t) = (t-T_{\mathrm{ref}})/P_{\mathrm{osc}}+\phi_{\mathrm{osc,ref}}$, $\phi_{\mathrm{rot}}(t) = (t-T_{\mathrm{ref}})/P_{\mathrm{rot}}+\phi_{\mathrm{rot,ref}}$, and $C_j$, which is a function of the varying radial velocity (see Sect.~\ref{section:sb1}). Analogously to Eq.~(\ref{eq:cosine_fit}), $\phi_{\mathrm{osc,ref}}$ and $\phi_{\mathrm{rot,ref}}$ are the phases at the fixed reference epoch $T_{\mathrm{ref}}$. Only a few of the available spectra\footnote{See Appendix~\ref{section:rvcurve} for an overview.} are of sufficient quality (high S/N, high $R$, suitable instrumental setup) for the spectral modeling of the pulsations. Ten strategically chosen metal lines are simultaneously fitted in these spectra, which cover five distinct epochs. Owing to the rugged $\chi^2$ landscape and because the model is extremely sensitive to tiny changes in several of the oscillation parameters, the numerical $\chi^2$ minimization normally fails to find the global best fit of this problem. However, given the simplicity of the model (only mono-periodic spheroidal modes are considered; atmospheric changes due to the pulsation are neglected) and the small number of epochs, the goal of the spectral modeling is merely to demonstrate that the observed line-profile variations are consistent with a pulsational signature rather than to determine the oscillation parameters with high fidelity. Hence, the search range for the oscillation period is limited to values close to the outcome of the light-curve analysis. For each combination of $l$ and $m$ up to $l=7$, the physical region of the multi-dimensional parameter space is explored by accepting only those models that give stellar radii or masses\footnote{The stellar radius $R_\star$ is derived from the identity $\varv\sin(i_{\mathrm{r}}) = \Omega R_\star \sin(i_{\mathrm{r}})$. The stellar mass $M$ then follows from $M = k^{(0)} (\omega^{(0)})^2 R_\star^3 / G$, where $G$ is the gravitational constant \citepads[see Eq.~(9) in]{1997A&AS..121..343S}.} in the range $5\,R_\sun \le R_\star \le 11\,R_\sun$ or $5\,M_\sun \le M \le 8\,M_\sun$, respectively. The results are summarized in Table~\ref{table:pulsation_modeling} and reveal that the spectral variations are best reproduced by a pulsational mode with $l=5$ and $m=1$. A comparison of this model with the observations is shown in Fig.~\ref{fig:pulsation_modeling_all}. Despite the simplicity of the applied theory, the overall match is very promising, which demonstrates that the line-profile distortions can be explained well by pulsations. 

Recalling Sect.~\ref{section:sb1} and Fig.~\ref{fig:velocity_shift}, it is worthwhile noting that the two spectra taken at MJD $53\,152.42$ and $53\,224.31$ have almost the same oscillation and rotation phase.

\begin{table*}
\caption{\label{table:pulsation_modeling}Parameters, $\chi^2$ statistics, and derived quantities of the slow-rotation model for all possible oscillation modes up to degree $l=7$.}
\tiny
\setlength{\tabcolsep}{0.083cm}
\renewcommand{\arraystretch}{1.03}
\begin{tabular}{rrrrrrrrrrrrrrrrrrr}
\hline\hline \\[-1em]
& & \multicolumn{8}{c}{Parameters} & & \multicolumn{8}{c}{Derived quantities} \\
\cline{3-10} \cline{12-19} \\[-1em]
$l$ & $m$ & $k^{(0)}$ & $P_{\mathrm{osc}} = 2\pi/\omega$ & $\phi_{\mathrm{osc,ref}}$ & $\phi_{\mathrm{rot,ref}}$ & $\Omega/\omega$ & $\varv\sin(i_{\mathrm{r}})$ & ${\langle \varv_\mathrm{v}^2 \rangle}{}^{1/2}$ & $i_{\mathrm{r}}$ & $\chi^2_{\mathrm{reduced}}$ & $a_{\mathrm{sph}}$ & $\omega^{(0)}$ & $\Omega/\omega^{(0)}$ & $\eta$ & $M$ & $R_\star$ & $P_{\mathrm{rot}} = 2\pi/\Omega$ & $\log(g\,\mathrm{(cm\,s^{-2})})$ \\ 
\cline{8-9} \\[-1em]
& & & (days) & & & & \multicolumn{2}{c}{($\mathrm{km\,s^{-1}}$)} & ($\degr$) & &  ($R_\sun$) & (days$^{-1}$) & & & ($M_\sun$) & ($R_\sun$) & (days) & (dex) \\
\hline
$5$ & $+1$ & $ 0.792$ & $ 1.3818$ & $0.4963$ & $0.5323$ & $0.0576$ & $ 16.07$ & $  1.96$ &  $44.2$ & $      4.3$ & $0.2688$ & $4.5429$ & $0.0577$ & $0.0042$ & $   7.3$ & $  10.9$ & $23.9801$ & $3.22$ \\ 
$6$ & $+1$ & $ 0.676$ & $ 1.3818$ & $0.5009$ & $0.5472$ & $0.0671$ & $ 16.28$ & $  2.15$ &  $39.7$ & $      5.3$ & $0.2938$ & $4.5435$ & $0.0672$ & $0.0067$ & $   5.3$ & $  10.4$ & $20.5795$ & $3.13$ \\ 
$7$ & $+1$ & $ 1.233$ & $ 1.3884$ & $0.5391$ & $0.5624$ & $0.0771$ & $ 15.27$ & $  1.29$ &  $36.7$ & $      5.4$ & $0.1774$ & $4.5224$ & $0.0771$ & $0.0048$ & $   6.4$ & $   9.1$ & $18.0159$ & $3.33$ \\ 
$4$ & $-1$ & $ 1.536$ & $ 1.3839$ & $0.7075$ & $0.4535$ & $0.0776$ & $ 16.07$ & $  1.53$ &  $40.8$ & $      6.0$ & $0.2099$ & $4.5492$ & $0.0774$ & $0.0039$ & $   7.1$ & $   8.7$ & $17.8413$ & $3.41$ \\ 
$3$ & $-1$ & $ 1.142$ & $ 1.3851$ & $0.1256$ & $0.6720$ & $0.0964$ & $ 18.16$ & $  1.48$ &  $35.5$ & $      6.1$ & $0.2030$ & $4.5545$ & $0.0960$ & $0.0081$ & $   5.6$ & $   8.9$ & $14.3632$ & $3.29$ \\ 
$1$ & $+1$ & $ 0.909$ & $ 1.3813$ & $0.9207$ & $0.4446$ & $0.0768$ & $ 18.03$ & $  3.94$ &  $36.4$ & $      6.1$ & $0.5389$ & $4.4614$ & $0.0783$ & $0.0067$ & $   7.8$ & $  10.8$ & $17.9848$ & $3.26$ \\ 
$7$ & $-3$ & $ 0.819$ & $ 1.3823$ & $0.1565$ & $0.4628$ & $0.0769$ & $ 19.41$ & $  1.37$ &  $43.1$ & $      6.1$ & $0.1873$ & $4.5549$ & $0.0767$ & $0.0072$ & $   5.9$ & $  10.1$ & $17.9838$ & $3.20$ \\ 
$7$ & $+2$ & $ 1.207$ & $ 1.3870$ & $0.6614$ & $0.5348$ & $0.0771$ & $ 15.80$ & $  1.12$ &  $35.1$ & $      6.2$ & $0.1534$ & $4.5238$ & $0.0772$ & $0.0049$ & $   7.9$ & $   9.8$ & $17.9975$ & $3.35$ \\ 
$5$ & $-2$ & $ 1.317$ & $ 1.3814$ & $0.1746$ & $0.5724$ & $0.0576$ & $ 14.91$ & $  1.76$ &  $50.7$ & $      6.3$ & $0.2407$ & $4.5571$ & $0.0575$ & $0.0025$ & $   7.1$ & $   9.1$ & $23.9694$ & $3.37$ \\ 
$5$ & $+2$ & $ 0.769$ & $ 1.3803$ & $0.6947$ & $0.5746$ & $0.0586$ & $ 16.46$ & $  2.01$ &  $44.3$ & $      6.3$ & $0.2747$ & $4.5431$ & $0.0587$ & $0.0045$ & $   7.1$ & $  11.0$ & $23.5749$ & $3.21$ \\ 
$6$ & $-4$ & $ 0.909$ & $ 1.3832$ & $0.7240$ & $0.5129$ & $0.0659$ & $ 19.42$ & $  1.24$ &  $48.8$ & $      6.5$ & $0.1699$ & $4.5567$ & $0.0657$ & $0.0048$ & $   7.9$ & $  10.7$ & $20.9742$ & $3.27$ \\ 
$4$ & $+1$ & $ 1.234$ & $ 1.3811$ & $0.9872$ & $0.5926$ & $0.0689$ & $ 16.22$ & $  1.96$ &  $42.0$ & $      6.6$ & $0.2685$ & $4.5415$ & $0.0691$ & $0.0039$ & $   7.6$ & $   9.6$ & $20.0320$ & $3.36$ \\ 
$4$ & $+2$ & $ 1.321$ & $ 1.3897$ & $0.7807$ & $0.6111$ & $0.0686$ & $ 15.06$ & $  1.38$ &  $44.4$ & $      6.6$ & $0.1900$ & $4.5059$ & $0.0688$ & $0.0036$ & $   5.9$ & $   8.6$ & $20.2708$ & $3.33$ \\ 
$2$ & $+0$ & $ 0.545$ & $ 1.3813$ & $0.8712$ & $0.4993$ & $0.0746$ & $ 16.78$ & $  3.76$ &  $34.1$ & $      6.6$ & $0.5147$ & $4.5487$ & $0.0746$ & $0.0102$ & $   5.0$ & $  11.0$ & $18.5110$ & $3.06$ \\ 
$2$ & $-2$ & $ 0.539$ & $ 1.3957$ & $0.5391$ & $0.3711$ & $0.0739$ & $ 19.12$ & $  2.85$ &  $40.4$ & $      6.7$ & $0.3944$ & $4.5571$ & $0.0730$ & $0.0099$ & $   5.1$ & $  11.0$ & $18.8904$ & $3.06$ \\ 
$4$ & $-3$ & $ 1.140$ & $ 1.3975$ & $0.7082$ & $0.6050$ & $0.0743$ & $ 17.75$ & $  0.98$ &  $49.8$ & $      6.8$ & $0.1363$ & $4.5211$ & $0.0739$ & $0.0048$ & $   5.1$ & $   8.6$ & $18.8158$ & $3.27$ \\ 
$6$ & $-3$ & $ 0.899$ & $ 1.3833$ & $0.0071$ & $0.5219$ & $0.0673$ & $ 18.87$ & $  1.36$ &  $49.5$ & $      6.9$ & $0.1862$ & $4.5532$ & $0.0672$ & $0.0050$ & $   6.5$ & $  10.1$ & $20.5470$ & $3.24$ \\ 
$6$ & $-6$ & $ 0.751$ & $ 1.3937$ & $0.4611$ & $0.4594$ & $0.0658$ & $ 19.11$ & $  1.28$ &  $48.5$ & $      6.9$ & $0.1760$ & $4.5295$ & $0.0655$ & $0.0057$ & $   6.4$ & $  10.7$ & $21.1718$ & $3.19$ \\ 
$6$ & $+0$ & $ 0.919$ & $ 1.3834$ & $0.3775$ & $0.5611$ & $0.0673$ & $ 16.23$ & $  2.37$ &  $38.4$ & $      6.9$ & $0.3254$ & $4.5420$ & $0.0673$ & $0.0049$ & $   7.7$ & $  10.6$ & $20.5663$ & $3.27$ \\ 
$4$ & $+4$ & $ 0.690$ & $ 1.3962$ & $0.7106$ & $0.2175$ & $0.0702$ & $ 18.47$ & $  1.33$ &  $42.5$ & $      6.9$ & $0.1844$ & $4.4687$ & $0.0707$ & $0.0072$ & $   5.8$ & $  10.7$ & $19.8864$ & $3.14$ \\ 
$5$ & $+5$ & $ 0.688$ & $ 1.3808$ & $0.2273$ & $0.4272$ & $0.0545$ & $ 19.47$ & $  0.58$ &  $73.9$ & $      6.9$ & $0.0792$ & $4.5297$ & $0.0547$ & $0.0044$ & $   5.0$ & $  10.1$ & $25.3432$ & $3.13$ \\ 
$2$ & $+2$ & $ 1.093$ & $ 1.3942$ & $0.8837$ & $0.5255$ & $0.0746$ & $ 18.81$ & $  1.88$ &  $42.9$ & $      7.0$ & $0.2594$ & $4.4506$ & $0.0755$ & $0.0052$ & $   7.8$ & $  10.2$ & $18.6906$ & $3.31$ \\ 
$7$ & $-2$ & $ 0.975$ & $ 1.3817$ & $0.4457$ & $0.4625$ & $0.0768$ & $ 17.76$ & $  1.21$ &  $44.3$ & $      7.0$ & $0.1662$ & $4.5537$ & $0.0767$ & $0.0060$ & $   5.1$ & $   9.0$ & $17.9843$ & $3.23$ \\ 
$5$ & $+4$ & $ 0.713$ & $ 1.3807$ & $0.2725$ & $0.5985$ & $0.0581$ & $ 18.93$ & $  0.95$ &  $54.2$ & $      7.0$ & $0.1295$ & $4.5330$ & $0.0583$ & $0.0048$ & $   6.6$ & $  11.0$ & $23.7633$ & $3.18$ \\ 
$7$ & $+0$ & $ 1.096$ & $ 1.3883$ & $0.3865$ & $0.5701$ & $0.0771$ & $ 17.02$ & $  1.29$ &  $38.3$ & $      7.0$ & $0.1774$ & $4.5257$ & $0.0771$ & $0.0054$ & $   7.1$ & $   9.8$ & $18.0165$ & $3.31$ \\ 
$6$ & $-5$ & $ 1.071$ & $ 1.3847$ & $0.4185$ & $0.5136$ & $0.0659$ & $ 19.04$ & $  0.92$ &  $52.1$ & $      7.1$ & $0.1260$ & $4.5554$ & $0.0656$ & $0.0040$ & $   7.6$ & $  10.0$ & $21.0169$ & $3.32$ \\ 
$4$ & $+0$ & $ 1.403$ & $ 1.3838$ & $0.9448$ & $0.4823$ & $0.0766$ & $ 16.52$ & $  1.92$ &  $40.1$ & $      7.1$ & $0.2627$ & $4.5405$ & $0.0766$ & $0.0042$ & $   7.6$ & $   9.2$ & $18.0720$ & $3.39$ \\ 
$5$ & $-1$ & $ 0.815$ & $ 1.3838$ & $0.2473$ & $0.6037$ & $0.0586$ & $ 17.20$ & $  1.96$ &  $46.9$ & $      7.1$ & $0.2684$ & $4.5449$ & $0.0585$ & $0.0042$ & $   7.6$ & $  11.0$ & $23.6274$ & $3.24$ \\ 
$6$ & $-2$ & $ 1.030$ & $ 1.3841$ & $0.2414$ & $0.5387$ & $0.0673$ & $ 16.95$ & $  1.49$ &  $48.2$ & $      7.1$ & $0.2044$ & $4.5469$ & $0.0672$ & $0.0044$ & $   5.7$ & $   9.2$ & $20.5577$ & $3.26$ \\ 
$2$ & $-1$ & $ 0.979$ & $ 1.3952$ & $0.7438$ & $0.3904$ & $0.0718$ & $ 19.42$ & $  2.71$ &  $52.6$ & $      7.3$ & $0.3745$ & $4.5304$ & $0.0714$ & $0.0052$ & $   5.7$ & $   9.4$ & $19.4333$ & $3.25$ \\ 
$5$ & $-4$ & $ 1.069$ & $ 1.3812$ & $0.4125$ & $0.3427$ & $0.0576$ & $ 19.35$ & $  1.08$ &  $78.0$ & $      7.4$ & $0.1474$ & $4.5665$ & $0.0574$ & $0.0031$ & $   6.2$ & $   9.4$ & $23.9732$ & $3.29$ \\ 
$6$ & $+3$ & $ 0.573$ & $ 1.3855$ & $0.2597$ & $0.5681$ & $0.0674$ & $ 17.07$ & $  2.10$ &  $39.4$ & $      7.4$ & $0.2876$ & $4.5240$ & $0.0676$ & $0.0080$ & $   5.2$ & $  10.9$ & $20.5540$ & $3.08$ \\ 
$3$ & $+1$ & $ 1.623$ & $ 1.3831$ & $0.6094$ & $0.5803$ & $0.0796$ & $ 16.88$ & $  1.19$ &  $49.1$ & $      7.5$ & $0.1624$ & $4.5277$ & $0.0798$ & $0.0039$ & $   5.1$ & $   7.7$ & $17.3795$ & $3.38$ \\ 
$5$ & $+0$ & $ 0.940$ & $ 1.3832$ & $0.3577$ & $0.5768$ & $0.0586$ & $ 16.82$ & $  1.78$ &  $53.8$ & $      7.5$ & $0.2442$ & $4.5426$ & $0.0586$ & $0.0036$ & $   6.1$ & $   9.7$ & $23.6226$ & $3.25$ \\ 
$6$ & $+6$ & $ 1.039$ & $ 1.3862$ & $0.9538$ & $0.9418$ & $0.0670$ & $ 18.04$ & $  1.32$ &  $54.6$ & $      7.5$ & $0.1815$ & $4.5111$ & $0.0673$ & $0.0044$ & $   5.3$ & $   9.0$ & $20.6847$ & $3.25$ \\ 
$7$ & $+6$ & $ 1.102$ & $ 1.3819$ & $0.3203$ & $0.6680$ & $0.0804$ & $ 19.89$ & $  1.05$ &  $49.9$ & $      7.5$ & $0.1441$ & $4.5272$ & $0.0807$ & $0.0059$ & $   5.3$ & $   8.8$ & $17.1895$ & $3.27$ \\ 
$5$ & $-3$ & $ 0.923$ & $ 1.3853$ & $0.2485$ & $0.4441$ & $0.0576$ & $ 19.21$ & $  1.30$ &  $58.9$ & $      7.5$ & $0.1790$ & $4.5487$ & $0.0574$ & $0.0036$ & $   7.9$ & $  10.7$ & $24.0647$ & $3.28$ \\ 
$7$ & $-1$ & $ 1.136$ & $ 1.3800$ & $0.2391$ & $0.4212$ & $0.0759$ & $ 16.77$ & $  0.98$ &  $44.3$ & $      7.5$ & $0.1342$ & $4.5560$ & $0.0758$ & $0.0051$ & $   5.2$ & $   8.6$ & $18.1923$ & $3.28$ \\ 
$4$ & $-2$ & $ 1.375$ & $ 1.3888$ & $0.5970$ & $0.6288$ & $0.0743$ & $ 17.42$ & $  1.50$ &  $53.1$ & $      7.6$ & $0.2058$ & $4.5410$ & $0.0740$ & $0.0040$ & $   5.0$ & $   8.0$ & $18.6890$ & $3.33$ \\ 
$7$ & $+7$ & $ 1.139$ & $ 1.3840$ & $0.3603$ & $0.8341$ & $0.0911$ & $ 19.86$ & $  1.13$ &  $40.9$ & $      7.6$ & $0.1545$ & $4.5139$ & $0.0916$ & $0.0074$ & $   6.0$ & $   9.1$ & $15.1984$ & $3.29$ \\ 
$4$ & $+3$ & $ 0.558$ & $ 1.3875$ & $0.4925$ & $0.6222$ & $0.0701$ & $ 18.01$ & $  1.29$ &  $39.8$ & $      7.7$ & $0.1777$ & $4.5047$ & $0.0705$ & $0.0089$ & $   5.1$ & $  11.0$ & $19.7801$ & $3.06$ \\ 
$7$ & $-5$ & $ 0.753$ & $ 1.3875$ & $0.0735$ & $0.3664$ & $0.0826$ & $ 19.26$ & $  1.54$ &  $35.8$ & $      7.9$ & $0.2115$ & $4.5453$ & $0.0823$ & $0.0090$ & $   6.9$ & $  10.9$ & $16.8068$ & $3.20$ \\ 
$6$ & $+5$ & $ 0.904$ & $ 1.3939$ & $0.0170$ & $0.5899$ & $0.0674$ & $ 18.82$ & $  1.04$ &  $53.7$ & $      7.9$ & $0.1436$ & $4.4894$ & $0.0677$ & $0.0051$ & $   5.4$ & $   9.5$ & $20.6841$ & $3.21$ \\ 
$1$ & $+0$ & $ 0.552$ & $ 1.3812$ & $0.9277$ & $0.5236$ & $0.0750$ & $ 17.69$ & $  3.44$ &  $36.0$ & $      7.9$ & $0.4712$ & $4.5492$ & $0.0750$ & $0.0102$ & $   5.1$ & $  11.0$ & $18.4167$ & $3.07$ \\ 
$7$ & $+3$ & $ 1.147$ & $ 1.3867$ & $0.1874$ & $0.5493$ & $0.0771$ & $ 16.04$ & $  1.12$ &  $35.0$ & $      8.0$ & $0.1532$ & $4.5217$ & $0.0772$ & $0.0052$ & $   7.9$ & $   9.9$ & $17.9944$ & $3.34$ \\ 
$3$ & $-2$ & $ 0.545$ & $ 1.3894$ & $0.1333$ & $0.5666$ & $0.0896$ & $ 17.92$ & $  1.96$ &  $30.0$ & $      8.0$ & $0.2698$ & $4.5561$ & $0.0889$ & $0.0145$ & $   5.1$ & $  11.0$ & $15.5149$ & $3.06$ \\ 
$7$ & $-4$ & $ 0.731$ & $ 1.3806$ & $0.1750$ & $0.3787$ & $0.0769$ & $ 19.51$ & $  1.39$ &  $39.5$ & $      8.1$ & $0.1902$ & $4.5637$ & $0.0767$ & $0.0080$ & $   6.7$ & $  10.9$ & $17.9570$ & $3.19$ \\ 
$7$ & $-7$ & $ 0.947$ & $ 1.3972$ & $0.9249$ & $0.3570$ & $0.0904$ & $ 19.58$ & $  1.46$ &  $40.5$ & $      8.1$ & $0.2021$ & $4.5224$ & $0.0898$ & $0.0085$ & $   5.2$ & $   9.2$ & $15.4635$ & $3.22$ \\ 
$7$ & $-6$ & $ 0.822$ & $ 1.3977$ & $0.9766$ & $0.3665$ & $0.0825$ & $ 19.81$ & $  1.27$ &  $38.7$ & $      8.2$ & $0.1762$ & $4.5152$ & $0.0821$ & $0.0082$ & $   6.8$ & $  10.6$ & $16.9496$ & $3.22$ \\ 
$5$ & $+3$ & $ 0.834$ & $ 1.3814$ & $0.5376$ & $0.5662$ & $0.0586$ & $ 17.86$ & $  1.98$ &  $56.2$ & $      8.3$ & $0.2712$ & $4.5350$ & $0.0587$ & $0.0041$ & $   5.9$ & $  10.0$ & $23.5851$ & $3.20$ \\ 
$6$ & $-1$ & $ 0.764$ & $ 1.3840$ & $0.3804$ & $0.5393$ & $0.0673$ & $ 17.66$ & $  1.59$ &  $42.7$ & $      8.5$ & $0.2173$ & $4.5436$ & $0.0672$ & $0.0059$ & $   6.4$ & $  10.6$ & $20.5764$ & $3.19$ \\ 
$3$ & $+0$ & $ 2.644$ & $ 1.3850$ & $0.3184$ & $0.6185$ & $0.0807$ & $ 17.05$ & $  0.61$ &  $56.3$ & $      8.6$ & $0.0831$ & $4.5365$ & $0.0807$ & $0.0025$ & $   6.2$ & $   6.9$ & $17.1702$ & $3.55$ \\ 
$2$ & $+1$ & $ 1.150$ & $ 1.3860$ & $0.9319$ & $0.5183$ & $0.0746$ & $ 16.70$ & $  1.68$ &  $38.0$ & $      8.7$ & $0.2309$ & $4.5051$ & $0.0751$ & $0.0049$ & $   7.9$ & $  10.0$ & $18.5693$ & $3.34$ \\ 
$1$ & $-1$ & $ 1.192$ & $ 1.3833$ & $0.7813$ & $0.5990$ & $0.0741$ & $ 18.35$ & $  2.13$ &  $44.4$ & $      9.0$ & $0.2921$ & $4.6263$ & $0.0727$ & $0.0044$ & $   7.9$ & $   9.7$ & $18.6732$ & $3.36$ \\ 
$3$ & $-3$ & $ 0.686$ & $ 1.3836$ & $0.8712$ & $0.4988$ & $0.0897$ & $ 17.83$ & $  1.69$ &  $32.5$ & $      9.0$ & $0.2319$ & $4.5921$ & $0.0887$ & $0.0115$ & $   5.1$ & $  10.1$ & $15.4178$ & $3.13$ \\ 
$5$ & $-5$ & $ 0.771$ & $ 1.3836$ & $0.4334$ & $0.3416$ & $0.0580$ & $ 17.42$ & $  0.71$ &  $53.4$ & $      9.1$ & $0.0979$ & $4.5631$ & $0.0577$ & $0.0043$ & $   5.9$ & $  10.2$ & $23.8619$ & $3.19$ \\ 
$7$ & $+5$ & $ 0.691$ & $ 1.3865$ & $0.4376$ & $0.6696$ & $0.0731$ & $ 19.06$ & $  1.32$ &  $42.0$ & $      9.4$ & $0.1815$ & $4.5168$ & $0.0733$ & $0.0078$ & $   5.8$ & $  10.7$ & $18.9664$ & $3.15$ \\ 
$6$ & $+4$ & $ 0.531$ & $ 1.3927$ & $0.2685$ & $0.5754$ & $0.0673$ & $ 17.34$ & $  1.75$ &  $39.8$ & $      9.5$ & $0.2420$ & $4.4970$ & $0.0675$ & $0.0086$ & $   5.0$ & $  11.1$ & $20.6960$ & $3.05$ \\ 
$0$ & $+0$ & $ 0.808$ & $ 1.3835$ & $0.7017$ & $0.7845$ & $0.0827$ & $ 17.52$ & $  2.50$ &  $32.4$ & $      9.5$ & $0.3420$ & $4.5416$ & $0.0827$ & $0.0085$ & $   7.1$ & $  10.8$ & $16.7291$ & $3.22$ \\ 
$6$ & $+2$ & $ 0.684$ & $ 1.3805$ & $0.4750$ & $0.5388$ & $0.0672$ & $ 15.54$ & $  1.61$ &  $36.2$ & $      9.5$ & $0.2199$ & $4.5440$ & $0.0673$ & $0.0066$ & $   5.8$ & $  10.7$ & $20.5323$ & $3.15$ \\ 
$4$ & $-4$ & $ 0.585$ & $ 1.3938$ & $0.5172$ & $0.6031$ & $0.0746$ & $ 18.14$ & $  1.18$ &  $38.2$ & $      9.7$ & $0.1632$ & $4.5416$ & $0.0741$ & $0.0094$ & $   5.2$ & $  10.8$ & $18.6808$ & $3.09$ \\ 
$3$ & $+3$ & $ 0.950$ & $ 1.3818$ & $0.5339$ & $0.5900$ & $0.0780$ & $ 16.88$ & $  0.85$ &  $34.3$ & $      9.9$ & $0.1165$ & $4.5028$ & $0.0787$ & $0.0065$ & $   7.6$ & $  10.5$ & $17.7242$ & $3.28$ \\ 
$7$ & $+4$ & $ 0.866$ & $ 1.3868$ & $0.3483$ & $0.5457$ & $0.0761$ & $ 17.47$ & $  1.36$ &  $40.4$ & $      9.9$ & $0.1863$ & $4.5183$ & $0.0763$ & $0.0067$ & $   5.5$ & $   9.7$ & $18.2232$ & $3.20$ \\ 
$3$ & $+2$ & $ 1.109$ & $ 1.3809$ & $0.7472$ & $0.6237$ & $0.0781$ & $ 16.62$ & $  0.48$ &  $41.8$ & $     10.2$ & $0.0655$ & $4.5205$ & $0.0786$ & $0.0056$ & $   5.1$ & $   8.7$ & $17.6865$ & $3.26$ \\ 
\hline
\end{tabular}
\tablefoot{The rugged $\chi^2$ landscape and the model's extreme sensitivity to tiny changes in several of the oscillation parameters, means that the numerical $\chi^2$ minimization normally fails to find the global best fit; the results shown here are probably just good fits instead of the best fit. The quantity $\eta$ is the ratio of the centrifugal force to the gravitational force at the equator of the star \citepads[see Eq.~(30) in]{1997A&AS..121..343S}. The table is sorted in ascending order by the reduced $\chi^2$.}
\end{table*}
\end{appendix}
\end{document}